\documentclass[12pt]{article}
\usepackage{graphicx}
\usepackage[page]{appendix}
\usepackage{amsmath}
\usepackage{amsthm}
\usepackage{slashbox}
\usepackage{txfonts}
\usepackage{caption}
\usepackage{subcaption}
\usepackage{multirow}
\usepackage{url}
\usepackage[authoryear]{natbib}
\usepackage{rotating}
\usepackage{floatpag}

\newcommand{\blind}{1}
\newtheorem{lemma}{Lemma}[section]
\newtheorem{theorem}[lemma]{Theorem}

\newtheorem*{theorem*}{Theorem}

\theoremstyle{definition}

\theoremstyle{remark}

\begin{document}

\def\spacingset#1{\renewcommand{\baselinestretch}%
{#1}\small\normalsize} \spacingset{1}

\if1\blind
{
  \title{\bf The Adequate Bootstrap}
  \author{Toby Kenney\thanks{
    The authors gratefully acknowledge financial support from \textit{NSERC}}\hspace{.2cm}\\
    Department of Mathematics and Statistics, Dalhousie University\\
    and \\
    Hong Gu \\
    Department of Mathematics and Statistics, Dalhousie University}
  \maketitle
} \fi

\if0\blind
{
  \bigskip
  \bigskip
  \bigskip
  \begin{center}
    {\LARGE\bf The Adequate Bootstrap}
\end{center}
  \medskip
} \fi

\begin{abstract}
There is a fundamental disconnect between what is tested in a model
adequacy test, and what we would like to test. The usual approach is
to test the null hypothesis ``Model M is the true model.'' However,
Model M is never the true model. A model might still be
useful even if we have enough data to reject it. In this paper, we
present a technique to assess the adequacy of a model from the
philosophical standpoint that we know the model is not true, but we
want to know if it is useful.

Our solution to this problem is to measure the parameter uncertainty
in our estimates caused by the model uncertainty. We use bootstrap
inference on samples of a smaller size, for which the model cannot be
rejected. We use a model adequacy test to choose a bootstrap size with
limited probability of rejecting the model and perform inference for
samples of this size based on a nonparametric bootstrap. Our
idea is that if we base our inference on a sample size
at which we do not reject the model, then we should be happy with this
inference, because we would have been confident in it if our original
dataset had been this size.
\end{abstract}

\noindent%
\vfill

\newpage
\spacingset{1.45} 

\section{Introduction}

Model adequacy testing is less ubiquitous than it ought to be. Any
analysis should be acompanied by model adequacy testing. However, in
practice, this is not always the case. There are several reasons for
this. One particular reason is the fundamental disconnect between what
is tested, and what we would like to test. The usual approach to
testing model adequacy is to set up an hypothesis test. The null
hypothesis is ``Model $\mathcal M$ is the true model.'' However, when
we consider the famous words of \citet{Box}: ``All models are
wrong. Some models are useful.''  we see the problem with this
approach. We already know that the null hypothesis is false, and our
model is wrong. What we want to know from our test is whether the
model is useful. 

This problem has been known for a long time, since at least
\citet{Berkson}. There has been previous work on the question of how
we can assess how closely the model fits the data under the philosophy
that we know the model does not exactly fit the data, but assume it is
a good enough approximation for our purposes. There have been methods
based on measuring the distance from the data distribution to the
model distribution. \citet{HodgesLehmann} were the first to suggest
such a testing approach, testing whether the data come from a
distribution near to the hypothesised
distirbution. \citet{RudasCloggLindsay} used the mixture distance
between the data and the model distribution, where the mixture
distance is defined as the smallest mixing proportion such that the
empircal distribution arises as a mixture of the model distribution
plus some mixing distribution (this was for discrete distributions
with finite support). Later papers \citet{Xi} and \citet{XiLindsay}
identified some problems with this approach, and \citet{Lindsay}
suggest using Kullback-Leibler divergence instead of mixture
distance. This improves the asymptotic properties of the
estimator. \citet{Davies} also argues for the importance of being able
to make better judgements as to whether a model is a good
approximation, rather than whether it is true, and suggests the
Kolmogorov distance, which is conceptually quite similar to the mixing
distance of \citet{RudasCloggLindsay}, but for continuous data.

A different approach by \citet{Lindsay}, which is more closely
related to the approach presented in this paper, is to measure the
quality of the approximation by the number of data points required to
reject the proposed model. They call this the ``credibility
index''. This is a natural measure, but it is often not a helpful
measure, because it is not related very closely to our intuitive
notions of a close approximation. The number of data points needed to
reject a model depends on both the model and the type of data
available. For example, individual data provides more information than
grouped data, so the credibility index will often be larger for
grouped data than individual data. Different distributions can also be
easier to reject. As a simple example, for a Bernoulli distribution,
it is much easier to reject $p=0.00001$ than to reject $p=0.5$. This
means that the credibility index has different meanings for different
distributions, and it is not easy to decide how good an approximation
a given credibility index indicates. 

The question of whether a model is useful is of course
context-dependent. It depends what we want to achieve. There are two
main reasons for using a model. The first is that the model is
suggested by our understanding of or hypotheses about the underlying
scientific process, and by means of measuring model adequacy, we hope
to learn how well our hypothesised process reflects the reality.  It
is very difficult to do this purely statistically from the data ---
there are often many mechanisms that can describe data, so deciding
how well data should match a model to support the hypothesis that the
model reflects a large proportion of the underlying process is more a
philosophical problem than statistical, and so is beyond the scope of
this paper. The second reason is that we are confident that the model
reflects at least some aspects of the underlying process, and we want
to estimate some associated parameters of this process.

In this second case, our model is useful if our parameter estimates
are close to the true values from this underlying process. We can
measure this by forming a confidence interval for the estimated
parameters which takes into account any uncertainty in the
appropriateness of the model. The adequate bootstrap approach proposed
here estimates this uncertainty in parameter estimates due to model
misspecification by performing inference based on a sample size small
enough that we cannot reject the model.

The intuitive justification for this approach is simple. If we had
originally taken a sample of this size, we would not have rejected the
model adequacy, so we would have been happy with the resulting
confidence interval. If we are confident in this inverval based on a
sample we might have collected, we should not lose this confidence
just because we know that if we took a larger sample, we would reject
the model adequacy. After all, whenever we perform a parametric
analysis, we know that by taking a larger sample, we would reject the
model adequacy; so if we are willing to accept parametric analysis,
then we are implicitly accepting this confidence interval.

Our objective here is to produce a confidence interval for a method
which is as general as possible, in the sense of applying to as many
possible sources of model inadequacy as possible. We cannot in general
be sure that our adequate bootstrap interval will always contain the
``true'' value with high probability for any particular situation,
because it is possible to contrive situations where the distribution
of the data has the same model as the ``true'' model, but with
different parameter values. For example, we could find a probability
distribution, such that if we contaminate a standard normal with this
distribution, the result is another normal distribution with mean 0
and variance, say 1.1. If our objective is to find an interval for the
parameters of the distribution before contamination, we cannot hope to
always do this in a data-driven way, since there is no way to know
that this data set is subject to contamination. However, this is an
extremely contrived situation, and what we can hope to achieve is that
for random sources of model inadequacy, our method has a high chance
to cover the ``true'' parameter values. For a large sample size, the
principal source of variability in parameter value estimates is
uncertainty over the model, so that for any particular situation, the
adequate bootstrap confidence interval will vary very little with the
particular sample taken, and more with the relation between the true
and data models.

The structure of the paper is as follows. In Section~\ref{METHOD}, we
describe the methodology in detail. In Section~\ref{ABS}, we provide
methods for calculating the adequate bootstrap size, one based on the
central limit theorem; the other more direct, to deal with cases such
as contamination with very heavy tails, where the adequate bootstrap
size is too small for the central limit theorem to apply. In
Section~\ref{THEORY}, we derive some general theory about how well
this procedure will work. We derive this mostly under likelihood ratio
tests, likelihood intervals and standard mild regularity
conditions. We anticipate that the results should hold under more
general conditions, but it can be difficult to analyse these
situations.  In Section~\ref{SIMULATIONS}, we compare this theoretical
performance with observed performance on simulated data for two common
examples where the data distribution differs from the ``true''
distribution. In Section~\ref{REALDATA}, we apply the adequate
bootstrap to three real data examples: one example from stock market
data, which is often analysed using a log-normal distribution; two
examples of power-law distributions, which are applied to a large
range of different data sets, in some cases with limited fit to the
data being studied. The first of these power-law examples examines a
wide range of data sets for which power laws have been suggested,
including data sets from \citet{PowerLaw} who were studying whether
power laws were appropriate to these data sets. Our conclusions differ
from theirs because we base our conclusions on goodness-of-fit, rather
than hypothesis testing, so in cases where the data set was large,
they would reject the power law model, even though it fit the data
remarkably well. We conclude in Section~\ref{CONCLUSION}.

\section{Method}\label{METHOD}

\subsection{General Overview}

The idea is to take bootstraps that are smaller than our sample, so
that the variance in the bootstraps is comparable to the variance
caused by uncertainty over the true model. We choose our bootstrap
size so that the probability of our bootstraps being found inadequate
by our model adequacy test (at a specified size) equals a specified
value. For convenience of analysis, we have taken this specified value
to be 0.5. This seems an intuitively reasonable way to define the
boundary between bootstraps for which the model is adequate and those
for which it is not. The size of the model adequacy test is a matter
of personal choice. If we have knowledge of the underlying process,
then we can tune this to make the coverage of our confidence interval
equal to a pre-specified value. However, if we have this knowledge of
the underlying process, we may be better served by incorporating it
into a more specialised technique for finding the confidence
interval. The adequate bootstrap is designed to be a general-purpose
method suitable for a range of types of model misspecification, so is
most appropriate when the exact sources of model misspecification are
unknown. We will discuss the method for estimating the adequate
bootstrap size later in this section. Once we have calculated the
adequate bootstrap size, we then perform our inference for a sample of
this size. In the case of a bootstrap, we would simply draw bootstrap
replicates of size $n$ from our sample, and take the appropriate
quantiles of our distribution of estimated parameters from the
bootstrap samples.

\subsubsection{With or Without Replacement}

\citet{Lindsay} argued for using a subsampling method, since using
a bootstrap with replacement will give a biassed estimate for the
rejection probability of a sample of size $n$ from the
distribution. However, in our case, we are using a bootstrap to obtain
our confidence interval, so we are more interested in the probability
of a bootstrap (with replacement) rejecting the model, and taking a
bootstrap with replacement (somewhat tautologically) gives an unbiassed
estimator of this probability. Using the subsampling approach would
only be appropriate if we planned to use a subsampling approach for
the inference. However, the relationship between subsample size and
inference properties is more complicated and does not provide an
accurate measurement of the variability due to model uncertainty in the
case where the adequate bootstrap size is a large proportion of the
sample size. 

\subsection{Calculating the Adequate Bootstrap Size}\label{ABS}

From a computational point of view, the main issue is to calculate the
adequate bootstrap size, namely the size at which we will reject 50\%
of model adequacy tests. We present two approaches here. Both methods
are based on attempting to estimate the power curve --- that is, the
power of the test to reject the model as a function of sample
size. The first is non-parametric, based on isotonic regression and
should work reliably in all cases. This is the method that has been
used in the simulations and real-data examples in this paper. The
second is based on the normal approximation used for the chi-square
test. It assumes that the central limit theorem will apply, in the
sense that the adequacy test statistic can be represented as an
average of adequacy values for each individual data point and that the
central limit theorem can be applied to this average. This parametric
approach is likely to work fairly well for the cases where the
difference between the data distribution and the true distribution
manifests itself in the form of a large number of observations being
slightly different from what the model predicts. It is likely to work
less well if the model adequacy test statistic is affected by a small
number of very influential observations (e.g. a small number of
extreme outliers). In this case, the model adequacy test is likely to
depend mainly on the number of these extreme outliers, which will not
follow a normal distribution.

\subsubsection{Nonparametric Approach}\label{CalculatingBSSize}

Our non-parametric approach is based on the assumption that the
probability of rejection should be expected to be an increasing
function of sample size. In some cases, this may not be strictly
accurate, but it should be close enough to find a good estimate for
the adequate bootstrap size. Suppose we have a sample of adequacy test
results from bootstraps of different sizes, $(n_i,I_i)$, where $n_i$
is the sample size and $I_i$ is an indicator variable which is 1 if
the $i$th bootstrap is rejected by the adequacy test. We want to
estimate the probability $p(n)$ that a bootstrap of size $n$ is
rejected. We have that $p(n)$ is an increasing function of $n$. This
problem is an isotonic regression. The maximum likelihood solution is
given in \citet{AyerBrunkEwingReidSilverman}.

\begin{theorem}[Ayer, Brunk, Ewing, Reid, Silverman, 1955]
The maximum likelihood estimate of $p(x)$ subject to the constraint
that $p(x)$ is a non-decreasing function of $x$ is given by
$$\hat{p}(x)=\max_{y\leqslant x}\min_{z\geqslant x} \frac{S_{[y,z]}}{T_{[y,z]}}$$
where $S_{[y,z]}$ is the number of rejected bootstraps for sample
sizes between $y$ and $z$ and $T_{[y,z]}$ is the total number of
bootstraps with sample sizes between $y$ and $z$.
\end{theorem}

Using this maximum likelihood estimate, we need to develop a strategy
to choose our sampling to maximise the accuracy with which we estimate
the value of $x$ for which $p(x)=\alpha$. This problem is outside the
scope of this paper, and could be a fruitful topic for future
research. For the present paper, we have used the following
proceedure:

\begin{enumerate}

\item Perform one bootstrap for each value of $n$. (For some larger
  datasets, we perform boostraps only for every 100th value of $n$ or less).

\item Estimate $\hat{p}(x)$, and solve $\hat{p}(x)=\alpha$.

\item For $k=1,\ldots,10$:

\begin{enumerate}

\item Perform another $10k$ bootstraps at each size within
  $\frac{100}{k}$ of the estimated solution.

\item re-estimate $\hat{p}(x)$ and solve $\hat{p}(x)=\alpha$.

\end{enumerate}

\end{enumerate}

\subsubsection{Parametric Approach}

\label{BootstrapSize}

Suppose we are given the data sample. We want to determine the
probability that a bootstrap sample of size $n$ rejects the model
adequacy test. Suppose we are performing a Pearson chi-square test
with $K$ classes. We expect that the solution for this case should be
a reasonable approximation for the general solution using other
tests. The test statistic is $\sum_{i=1}^K \frac{(s_i-np_i)^2}{np_i}$,
where $s_i$ is the number of observations in the bootstrap sample that
lie in interval $i$, and $p_i$ is the probability that an observation
lies in interval $i$ given the parametric distribution. To simplify
things, assume we use the estimated parameters from the whole data,
rather than the bootstrap sample. This means that the number of
degrees of freedom should be $K-1$. This is slightly incorrect if $n$
is close to $N$, but if $n\ll N$, then it should be fairly accurate.

Now let $r_i$ be the empirical probability of class $i$ for the whole
data. That is, the number of points in class $i$ divided by the total
sample size $N$. Then $s_i$ follows a binomial distribution with
parameters $n$ and $r_i$. We approximate this as a normal distribution
with mean $nr_i$ and variance $nr_i(1-r_i)$. The covariance of $s_i$
and $s_j$ is $-nr_ir_j$.  The test statistic is therefore

\begin{align*}
X^2&=\sum_{i=1}^K \frac{(s_i-nr_i+nr_i-np_i)^2}{np_i}=\sum_{i=1}^K
\frac{r_i}{p_i}\left(\frac{(s_i-nr_i+nr_i-np_i)^2}{nr_i}\right)\\
&=\sum_{i=1}^K
\left(1+\frac{r_i-p_i}{p_i}\right)\left(\frac{(s_i-nr_i+nr_i-np_i)^2}{nr_i}\right)
\end{align*}

Let $E_i=s_i-nr_i$, $d_i=r_i-p_i$ and
$T_i=\frac{(E_i+nd_i)}{\sqrt{nr_i}}$. We have that $T_i$ are
approximately normally distributed with mean
$\frac{nd_i}{\sqrt{nr_i}}$ and variance $1-r_i$. Then
$T=(T_1,\ldots,T_K)$ follows a multivariate normal distribution, with
the covariance of $T_i$ and $T_j$ being $-\sqrt{r_ir_j}$. Let
$\Sigma_{ij}=\left\{\begin{array}{ll}1-r_i & \textrm{if
}i=j\\-\sqrt{r_ir_j}&\textrm{if }i\ne j\end{array}\right.$ be the
covariance matrix of $T$. If we let $v$ be a column vector with $i$th
entry $v_i=\sqrt{r_i}$, then we have $\Sigma=I-vv^T$, so for any
vector $w$, $\Sigma w=w-v(v^Tw)$; so if $w$ is orthogonal to $v$, then
we have that $\Sigma w=w$, so the nonzero eigenvalues of $\Sigma$ are
all 1.  Now if we let $A$ be a matrix of orthogonal unit nonzero
eigenvectors of $\Sigma$. We can write $T=\mu+AZ$ where $Z$ is a
vector of independent standard normal distributions. We also know that
$A^TA=I$. This gives
\begin{align*}
\sum_{i=1}^K T_i{}^2&=T^TT\\
&=\left(\mu+AZ\right)^T\left(\mu+AZ\right)\\
&=\mu^T\mu+2\mu^TAZ+Z^TA^TAZ\\
&=\mu^T\mu+2\mu^TAZ+Z^TZ\\
&=\left(A^T\mu+Z\right)^T\left(A^T\mu+Z\right)\\
\end{align*}
is a non-central chi-squared distribution with $K-1$ degrees of freedom. 
The Pearson chi-squared statistic is 
$$X^2=\sum_{i=1}^K \left(1+\frac{d_i}{p_i}\right)T_i{}^2$$ If our
model is approximately correct, so that we can neglect the
$\frac{d_i}{p_i}$ in this formula, then $X^2\approx\sum_{i=1}^K
T_i{}^2$, which approximately follows a non-central chi-square
distribution with $\lambda$ equal to $\sum_{i=1}^K
\frac{(nd_i)^2}{np_i}$, which is $\frac{n}{N}$ times the chi-square
statistic we obtain for the whole data set, and with $K-1$ degrees of
freedom. We want to know the probability that $X^2$ is sufficient to
reject the model adequacy test. That is, we want to know the
probability that this statistic is larger than the critical value for
a (central) chi-square distribution with $K-1$ degrees of
freedom. Since we set our cuttoff at a 50\% chance of rejection, we
are aiming to select $\lambda$ so that the median of this non-central
chi-squared distribution is the critical value of the central
chi-square distribution. For fixed $K$, we can find required value of
$\lambda$ by simulation or numerical integration. We can then
immediately estimate the adequate bootstrap size from the statistic
for the overall data set.

\section{Coverage Probability for Adequate Bootstrap Interval}\label{THEORY}

We derive the theory about coverage probability for the adequate
bootstrap interval. We cannot get good results regarding the coverage
probability under reasonable assumptions of the sort that are possible
for confidence intervals under an assumed parametric model. The reason
is that for the usual confidence interval, the model is assumed known,
and it is only the sample size that affects the distribution of the
confidence interval. For the adequate bootstrap, the coverage depends
on the data model and the true model. For a fixed choice of data model
and true model, and a large enough sample size, the adequate bootstrap
interval will either always cover the true parameter values, or always
fail to cover it (assuming we take enough bootstrap samples). The
point is that for a large sample size, the empirical distribution is
very close to the data distribution, so the distribution of bootstrap
samples of size $n$ is very close to the distribution of samples of
size $n$ from the data distribution. The coverage here comes from
assumptions about how the data model (viewed as a random variable)
might differ from the true model.

We use likelihood ratio tests and likelihood-based confidence
intervals to develop our coverage probability. For most tests, with a
fixed sample size, we can approximate the test as a likelihood ratio
test by carefully choosing the space of alternative models. Therefore,
the results in this section should give us a good idea of what we
should expect in general. In fact, when a likelihood ratio test is
used to assess model adequacy, it is not usually appropriate to use
the adequate bootstrap, because a likelihood ratio test not only
assesses the adequacy of the model, but if the model is deemed
inadequate, the likelihood ratio test often provides a better
model. If there is a better model, which is adequate, it will often be
preferable to use the better model for the analysis, rather than to
use the adequate bootstrap for inference. The use of adequate
bootstrap is for cases when there is no obvious candidate better
model. In these cases, we only know that the current model does not
describe the data well, but do not know how it should be changed to
better fit the data. In some cases, the alternative model in the
likelihood ratio test might not be considered better, for example, if
it is a saturated model. In this case the use of the adequate
bootstrap with a likelihood ratio test might be appropriate.

For this section, we will use the following notation: There is a
``true'' model, which comes from some parametric model ${\mathcal P}$,
with some parameter values $\theta$. There is a data model, which
comes from a larger parametric model ${\mathcal M}$ with more
parameters, and with parameter values $\phi$. We will use the notation
${\mathcal M}(\phi)$ to represent the model from family ${\mathcal M}$
with parameters $\phi$. The parametric family ${\mathcal P}$ is nested
within ${\mathcal M}$, and we define $\xi$ so that ${\mathcal
  M}(\xi)={\mathcal P}(\theta)$. Let the best approximation to this
data distribution from the family ${\mathcal P}$ be ${\mathcal
  P}(\theta')={\mathcal M}(\xi')$. We will let $m$ be the number of
parameters in the family $\mathcal P$ and $m+k$ be the number of
parameters in the family $\mathcal M$.

We define
\begin{align*}
A^2(\phi)&=2{\mathbb E}_{X\sim {\mathcal M}(\phi)}\left(l(X;\phi)-l(X;\xi')\right)\\
B^2(\phi)&=2{\mathbb E}_{X\sim {\mathcal M}(\phi)}\left(l(X;\xi')-l(X;\xi)\right)\\
\end{align*}
so $A^2$ is twice the KL divergence from ${\mathcal M}(\phi)$ to
${\mathcal P}(\theta')={\mathcal M}(\xi')$. When ${\mathcal
  M}(\phi)\approx{\mathcal P}(\theta')$, $B^2$ will approximate twice
the KL divergence from ${\mathcal P}(\theta')$ to ${\mathcal
  P}(\theta)$. If we have a sample of size $n$, then the expected
value of the log-likelihood ratio statistic for the model goodness-of-fit
test is $nA^2$. For large $n$, the central limit theorem says that
this log-likelihood ratio statistic will be approximately normal, so
the median will also be $nA^2$. Therefore, the probability of
rejecting the model adequacy test will be 0.5 when $nA^2$ is equal to
the critical value of the likelihood ratio test. If we let $c_k$ be
this critical value, then we get that the adequate bootstrap size is
$\frac{c_k}{A^2}$.

Suppose our confidence interval is closely approximated by an expected
log-likelihood interval: that is, an interval of the form ${\mathbb
  E}_{(X_1,\ldots,X_n)\sim {\mathcal
    M}(\phi)}\left(l(X_1,\ldots,X_n;\xi)\right)\geqslant l_c$ for some
value $l_c$. Standard asymptotic theory will give that this is a good
approximation for the adequate bootstrap interval in a range of cases
when the adequate bootstrap size is large enough. If the adequate
bootstrap size is too small, this approximation may not be valid. The
expected difference in log-likelihood between $\theta$ and $\theta'$
will be $nB^2$. Therefore, the interval will contain the true value
$\theta$ provided $nB^2<c_m$. Substituting the value of $n$ used
above, we get that the adequate confidence interval covers the true
value whenever $\frac{c_k B^2}{A^2}<c_m$ or equivalently
$\frac{B^2}{A^2}<\frac{c_m}{c_k}$.

As mentioned earlier, the coverage is taken over a distribution for
the parameter values $\phi$, relative to $\xi$. We assume the original
sample size is large enough that the effect of the particular data
sample is small, and what affects coverage for fixed $\xi$ is just the
parameter values $\phi$, which we will now view as a random
variable. To indicate that these parameter values are random, we will
denote the random parameter values $\Phi$.

\subsection{General Formula}\label{TCKH}

If we assume that the distortion is small enough that the KL
divergence is well approximated by a quadratic approximation, then if
the distortion from the ``true'' model is a vector $x=\phi-\xi$, then
the KL divergence from the distorted model to the true model is
$\frac{1}{2}x^THx$ where $H$ is the Fisher information matrix, given
by
$$H_{ij}=-{\mathbb E}_{X\sim {\mathcal
    M}(\xi)}\left(\left.\frac{\partial^2
  l(X;\phi)}{\partial\phi_i\partial\phi_j}\right|_{\phi=\xi}\right)$$
The best-fitting distribution from the model distribution is then a
distortion $t=\xi'-\xi$ from the true model, where $t$ only takes
non-zero values in the parameters in the fitted model $\mathcal
P$. The KL divergence from the distorted model to the fitted model is
therefore
$\frac{1}{2}(x-t)^TH(x-t)=\frac{1}{2}\left(x^THx-2t^THx+t^THt\right)$. We
reparametrise $\mathcal M$ in such a way that changes to the first $m$
parameters remain within the family $\mathcal P$. That is, we rewrite
$x=(x_1,x_2)$, where $x_1$ represents distortion within $\mathcal P$
and $x_2$ represents distortion in the other parameters. Let the
Fisher information matrix with respect to this parametrisation be
$$H=\left(\begin{array}{ll}H_{11}&H_{12}\\H_{21}&H_{22}\end{array}\right)$$
The KL divergence is therefore
$$\frac{1}{2}\left(x_1{}^TH_{11}x_1+2x_1{}^TH_{12}x_2+x_2{}^TH_{22}x_2-2t^T(H_{11}x_1+H_{12}x_2)+t^TH_{11}t\right)$$
Setting the derivatives with respect to $t$ to zero, we get
$$t=H_{11}{}^{-1}(H_{11}x_1+H_{12}x_2)=x_1+H_{11}^{-1}H_{12}x_2$$
This gives us:
$$x-t=(-H_{11}^{-1}H_{12}x_2,x_2)$$
Substituting this into the formula for KL-divergence gives 
\begin{align*}
A^2&=(x-t)^TH(x-t)\\
&=\left(x_2{}^TH_{22}x_2+x_2{}^T(H_{21}H_{11}{}^{-1}H_{12})x_2-2x_2{}^T(H_{21}H_{11}{}^{-1}H_{12})x_2\right)\\
&=\left(x_2{}^TH_{22}x_2-x_2{}^T(H_{21}H_{11}{}^{-1}H_{12})x_2\right)\\
&=\left(x_2{}^T(H_{22}-H_{21}H_{11}{}^{-1}H_{12})x_2\right)\\
\end{align*}
Therefore, the adequate bootstrap size is $\frac{c_{k}}{x_2{}^T(H_{22}-H_{21}H_{11}{}^{-1}H_{12})x_2}$.
and
\begin{align*}
B^2&=t^TH_{11}t\\
&=(x_1+H_{11}^{-1}H_{12}x_2)^TH_{11}(x_1+H_{11}^{-1}H_{12}x_2)\\
&=\left(x_1{}^TH_{11}x_1+2x_1^TH_{12}x_2+x_2{}^T(H_{21}H_{11}{}^{-1}H_{12})x_2\right)\\
\end{align*}

We can then use this approach to estimate the probability
\begin{equation}\label{EqCovProb}
P\left(\frac{B^2}{A^2}\leqslant\frac{c_m}{c_k}\right)=P\bigg(c_m\left(x_2{}^T(H_{22}-H_{21}H_{11}{}^{-1}H_{12})x_2\right)\geqslant
c_k \left(x_1{}^TH_{11}x_1+2x_1^TH_{12}x_2+x_2{}^T(H_{21}H_{11}{}^{-1}H_{12})x_2\right)\bigg)\end{equation}

We demonstrate this technique for calculating coverage in our second
simulation study.

\subsection{Distortion following Inverse Fisher Information Matrix}\label{FISHERDISTORTION}

The previous section gives a formula for the coverage of the adequate
bootstrap interval. However, the difficult part in evaluating this
coverage is the distribution of $\phi$ (or equivalently $x=\phi-\xi$
in Formula~(\ref{EqCovProb})). If we know the particular distribution
of $\phi$, we can model the distortion directly, rather than use our
adequate bootstrap. To apply this, we will therefore need to choose a
baseline assumption for the distribution of $\Phi$. The most natural
assumption is that $\Phi$ is normally distributed with mean
$\xi$ and variance the inverse of the Fisher information matrix at
$\xi$ (the matrix $H$ in Section~\ref{TCKH}). The Fisher information
matrix gives the natural coordinate system for the parametric family
$\mathcal M$ in the neighbourhood of $\xi$, so this is as natural a
baseline assumption as any we might make.

Assuming the Fisher information matrix is approximately constant on
the region we are interested in, we can reparametrise $\mathcal M$ so
that the Fisher information matrix is the identity, and $\phi$
therefore follows a standard normal distribution. In this case,
substituting the identity matrix for $H$ in Equation~(\ref{EqCovProb}) gives 
$$P\left(\frac{B^2}{A^2}\leqslant\frac{c_m}{c_k}\right)=P\bigg(c_m\left(x_2{}^Tx_2\right)\geqslant
c_k \left(x_1{}^Tx_1\right)\bigg)$$
where $x_1$ and $x_2$ follow independent standard multivariate normal
distributions in $m$ and $k$ dimensions respectively. The products
$x_1{}^Tx_1$ and $x_2{}^Tx_2$ therefore follow independent chi-square
distributions with $m$ and $k$ degrees of freedom respectively. We
therefore get that $\frac{kx_1{}^Tx_1}{mx_2{}^Tx_2}$ follows an $F$
distribution with parameters $m$ and $k$, and we have 
$$P\left(\frac{B^2}{A^2}\leqslant\frac{c_m}{c_k}\right)=P\left(\frac{kx_1{}^Tx_1}{mx_2{}^Tx_2}\leqslant
\frac{kc_m}{mc_k}\right)$$ 
That is, probability of covering the true values is the probability
that an $F$ distribution with parameters $m$ and $k$ is at most
$\frac{kc_m}{mc_k}$. We summarise these theoretical coverages for
small values of $k$ and $m$ in Table~\ref{TheoreticalCoverage} in Appendix~\ref{AppendixTables}.

Note from Table~\ref{TheoreticalCoverage}, that the coverage increases
with $k$ and decreases with $m$. The intuitive explanation for this is
that we are assuming that the distortion $\phi-\xi$ is in a uniformly
random direction. If it is in a direction close to the $m-dimensional$
subspace $\mathcal P$, then it represents a large change in the
parameters $\theta'$, but a fairly good fit, so will result in lower
coverage. The probability of being close to the subspace $\mathcal P$,
is an increasing function of $m$ and a decreasing function of $k$.

For more general goodness-of-fit tests such as Anderson-Darling or
Kolmogorov-Smirnov, we can approximate the behaviour as that of a
likelihood ratio test with very large degrees of freedom. If we let
$k\rightarrow\infty$, then the chi-square distribution is
approximately normal with mean $k$ and variance $2k$. If we let $c$ be
the critical value of a standard normal, then we have
$c_k=k+c\sqrt{2k}$. The limit as $k\rightarrow\infty$ of
$k\frac{B^2}{A^2}$ follows a chi-squared distribution with $m$ degrees
of freedom, so the coverage probability is the probability that a
chi-squared distribution with $m$ degrees of freedom is smaller than
$\frac{kc_m}{c_k}$, and since $\frac{k}{c_k}\rightarrow 1$, this gives
us the probability that the chi-squared random variable with $m$
degrees of freedom is smaller than $c_m$, which is exactly the
coverage we were aiming for.

It is worth noting that in this case, the coverage does not appear to
depend upon either the size of the model adequacy test, or the choice
of 50\% power in the definition of adequate bootstrap size. The
explanation of this is that this approximation as a high-dimensional
likelihood ratio test is most appropriate when the adequate bootstrap
size is large. In this case, for the large bootstrap size, there is
relatively less variability in the bootstrap samples, so the power
curve is relatively steeper in this case, meaning that the relative
change in adequate bootstrap size caused by changing these values is
small, meaning they have little effect on the adequate bootstrap
interval.

\subsection{Finite Sample Size}\label{FiniteSample}

The previous theory is assuming that the original sample size is
extremely large, so that the empirical distribution and the data
distribution are approximately equal. We suppose now that the data
sample size is $N$. \citet{Lindsay} point out that taking a
bootstrap with replacement and taking a subsample give different
power to the model adequacy test. For our purposes, this is not an
important distinction --- the bootstrap with replacement and the
subsample both give pseudo-samples that we can use for inference. We
calculate the size at which these pseudo-samples start to become
significantly different from samples from the model distribution,
and use samples of this size for our inference. It does not matter
that this size is different for the bootstrap with replacement and the
subsample. What matters is that we use the appropriate size for the
samples we are taking.

For a finite sample of size $N$, we can use the results from
Sections~\ref{TCKH} and~\ref{FISHERDISTORTION} on the empirical
distribution, rather than the data distribution. That is, we replace
$\phi$ by parameters corresponding to the empirical distribution. (We
will assume that $\mathcal M$ is a large enough model space that it
can include the empirical distribution.) The effect of this is that
the parameters $\phi$ for the empirical distribution are usually
further from $\xi$ than when we took the data distribution (since
there is additional noise). However, since it is the direction of
$\phi-\xi$ that is important for the adequate bootstrap coverage, this
is not important. Provided our sample size is large enough that the
empirical distribution is not too far from the data distribution, and
the empirical distribution does not cause some bias in this direction,
our results on coverage should still hold for this situation.

This demonstrates an advantage of using bootstrap samples, rather than
subsamples for deriving the theory of adequate bootstraps. If we take
a subsampling approach then we cannot consider our subsample to be
independent samples from the empirical distribution, since they are
drawn without replacement. For the case where the adequate bootstap
size is close to the sample size, this can have a sizeable effect on the
coverage probability.


\section{Simulations}\label{SIMULATIONS}

\subsection{Contaminated Normal}

%
%

Contamination is a good example where there is a well-defined ``true''
model with scientifically meaningful parameter values, which does not
match the data distribution. For this example, we can examine the
coverage of the adequate bootstrap confidence interval.

\subsubsection{Simulation Design}

We set the main distribution as a standard normal distribution. We
assume the standard deviation is known, and consider the estimation of
the mean. Since we are using a bootstrap-based method, there is no
great additional difficulty to using an estimated variance. The
computations will be slightly longer because we have to estimate the
variance in addition to the mean for each adequacy test. We consider
two important factors --- contaminating distribution and contamination
proportion. Since for any particular contaminating distribution and
proportion, the coverage probability of the adequate bootstrap
confidence interval should tend to either 0 or 1 for large sample
size, we let the parameters of the contaminating distribution be
random variables. In particular, we let the contaminating standard
deviation take a fixed value $\sigma$, but we let the contaminating
mean follow a normal distribution with mean 3 and standard deviation
$\tau$. We will refer to the distribution of the contaminating
parameters as the contaminating hyperdistribution.  We will do two
simulations, one to study the effect of the contaminating
hyperdistribution, and the other to study the effect of contamination
proportion.

In the first simulation, we fix the contamination proportion at 2\%,
and we use three contaminating hyperdistributions: $\sigma=1,\tau=6$,
$\sigma=3,\tau=4$, and $\sigma=8,\tau=1$. For each contaminating
hyperdistribution, we simulate 1000 contaminating distributions, and
for each contaminating distribution, we simulate a data set with a
total of 20000 data points from the contaminated distribution, and
apply the adequate bootstrap with the size of the adequacy test set to
5\% and bootstrap coverage set to 95\%, to obtain a confidence
interval for the mean $\mu$ of the main distribution.

In the second simulation, we fix the contaminating hyperdistribution
parameters as $\sigma=3,\tau=4$, and simulate under five different
contaminating proportions: 0.0001, 0.0005, 0.001, 0.02 and 0.05. For
each contamintation proportion, we simulate 1000 contaminating
distribution parameters from the hyperdistribution, and for each
contaminating distribution, we simulate 20000 data points, and perform
an adequate bootstrap to obtain a confidence interval, 
with the size of the adequacy test set to 5\%  and bootstrap coverage set to 95\%.

\subsubsection{Results}

Table~\ref{ContMean} gives the coverage of the adequate bootstrap
interval for the mean of the normal distribution for a range of
parameter values for the contaminating hyperdistribution, with
contamination percentage set to 2\%, and compares this to the coverage
of a standard bootstrap ignoring the model inadequacy issue. We see
that the $\sigma=1,\tau=6$ case has relatively lower coverage, since
these hyperparameter values give a relatively high probability that the
contaminated data look very similar to a normal
distribution. In these cases, the model adequacy test has more
difficulty rejecting the normal hypothesis, so the adequate bootstrap
size can be large, but the contamination changes the mean. In the
other scenarios, the contaminated distributions are far less similar
to a normal distribution, so the adequate bootstrap size is smaller,
and therefore achieves good coverage.

\begin{table}[htbp]
\caption{Number of Confidence Intervals containing the true
  mean out of 1000 simulations, contamination proportion=0.02}\label{ContMean}

\hfil\begin{tabular}{lrr}
\hline
Parameters& Adequate Bootstrap & Standard Bootstrap \\
\hline
$\sigma=1,\tau=6$ &  880 &  86   \\
$\sigma=3,\tau=4$ & 1000 & 131   \\
$\sigma=8,\tau=1$ & 1000 & 146   \\  
\hline
\end{tabular}

\end{table}

Table~\ref{ContProb} shows the coverage of the adequate bootstrap and
the standard bootstrap for varying contamination proportion, where the
mean of the contaminating distribution follows a normal distribution
with mean 3 and standard deviation 4. Table~\ref{AdBootSize} shows the
median adequate bootstrap size for the 1000 simulations in each
scenario. (We use the median because the adequate bootstrap size can
have a heavy-tailed distribution, so the median seems a better
indicator of a typical situation). For small contamination
probability, the data distribution is very close to the true
distribution, so the standard bootstrap has good coverage, and the
adequate bootstrap size is usually the sample size. This means that
the adequate bootstrap interval has not become substantially wider
than necessary in this case. As the contamination proportion
increases, the data becomes less normal, and the contamination affects
the mean so badly it is outside the standard bootstrap confidence
interval. As this happens, it becomes easier to reject the normal
distribution, so the adequate bootstrap size decreases. This makes the
adequate bootstrap confidence interval wider, so it retains good
coverage.

\begin{table}[htbp]
\caption{Number of Confidence Intervals containing the true
  mean out of 1000 simulations, $\sigma=3,\tau=4$}\label{ContProb}

\hfil\begin{tabular}{lrr}
\hline
Contamination& Adequate & Standard \\
Proportion & Bootstrap & Bootstrap\\
\hline
0.0001  &  948 & 940 \\
0.0005  &  941 & 922 \\
0.001   &  936 & 882 \\
0.02    & 1000 & 131 \\
0.05    & 1000 &  52 \\
\hline
\end{tabular}
  
\end{table}

\begin{table}[htbp] 
\caption{Median adequate bootstrap size and confidence interval width}\label{AdBootSize}

\hfil\begin{tabular}{lrr}
\hline
Contamination & Median Adequate & Median
Confidence \\
Proportion& Bootstrap Size & Interval Width\\
\hline
0.0001  & 20000 & 0.0279 \\
0.0005  & 20000 & 0.0281 \\
0.001   & 19814 & 0.0285 \\
0.02    & 101 & 0.5738 \\
0.05    & 48 & 0.7930\\
\hline
\end{tabular}
  
\end{table}

\subsection{Sampling Bias}

In this case, we consider a normal distribution with mean known to be
zero, but unknown variance $\sigma^2$. We imagine that this represents
the underlying distribution, but that the data are subject to some sampling
bias. That is, each data point $x$ has probability $g(x)$ of being
included in the sample.  Since the resulting data distribution only
depends on the relative sizes of the $g(x)$, it will be convenient to
rescale the $g(x)$, so that the values of $g(x)$ used may not actually
be probabilities. For simplicity, we let the sampling bias be a
stepwise function given by $g(x)=p_i$ whenever $c_{i-1}<x\leqslant
c_{i}$ where $-\infty=c_0<c_1<\cdots<c_{J-1}<c_J=\infty$ are the
boundary points. For simplicity we assume the $c_i$ are known and only
the $p_i$ need to be estimated. For the simulation, we set $c_i$ equal
to the $100\frac{i}{J}$th percentile of the ``true'' distribution, and
simulate $p_i$ following a log-normal distribution with $\mu=0$ and a
specified value of $\sigma$, which we will denote $\tau$ to
distinguish it from the parameter of the ``true'' normal
distribution. Our model adequacy test is a likelihood ratio test
between the true distribution and the distribution with sampling bias
with the $c_i$ fixed and the $p_i$ estimated (scaling the $p_i$ makes
no difference, so we choose the scale conveniently).

We consider a range of scenarios, varying both the number of steps in
the sampling probability function, and the parameter $\tau$ for the
log-normal distribution of the $p_i$. We consider three values for the
number of steps $J$: 3, 5 and 8. For each of these, we consider two
values for $\tau$: 0.2 and 0.5. We also consider the null scenario
where the $\tau=0$, so all $p_i$'s are 1, and the ``true''
distribution is the same as the data distribution. We simulate 1000
sets of values for $p_i$'s under each scenario, and for each set of
$p_i$ values, we simulate 20000 data points, and perform an adequate
bootstrap on them to estimate the unknown parameter $\sigma$ of the
``true'' distribution.

The model adequacy test used in this example is a likelihood ratio
test against the data distribution. This is unrealistic, because in
practical cases where we might use the adequate bootstrap, we would
not know the data distribution, so could not use a likelihood ratio
test against it. However, the likelihood ratio test allows us to see
the quality of the theoretical estimates we made about the coverage
that could be achieved.

\subsubsection{Estimating the MLEs}

The log-likelihood for this model is given by the sum of
log-likelihood of each observation conditional on a randomly sampled
point begin included in the data set. That is,
$$l(x)-\sum \frac{x_i{}^2}{2\sigma^2}-n\log(\sigma)+\sum_{i=1}^{J}
n_i\log(p_i)-n\log(\sum_{i=1}^{J} T_ip_i)$$
where:
\begin{itemize}
\item $n$ is the sample size

\item $n_i$ is the number of observations in the interval
  $(c_{i-1},c_i]$.

\item $T_i$ is the probability that a random sampling point from a
  normal distribution with mean 0 and variance the current estimate of
  $\sigma$ lies in $(c_{i-1},c_i]$. That is
    $T_i=\Phi\left(\frac{c_i}{\sigma}\right)-\Phi\left(\frac{c_{i-1}}{\sigma}\right)$.
\end{itemize}

The first two terms are the sums of log-likelihood of the points $x_i$ under the
``true'' normal distribution. The third term is the sum of the
log-likelihoods of the points $x_i$ being included in the sample. The
final term is $-n$ times the log-likelihood of a random point following
the estimated normal distribution being included in the sample.
For fixed $\sigma$, it is easy to see that the likelihood is maximized
by $p_i=\frac{n_i}{T_i}$ (rescaling all $p_i$ by the same factor
doesn't change the likelihood, so we have chosen a convenient
scaling). We can substitute this into our likelihood function, to get
a univariate function of $\sigma$. We can then use Newton's method to
find the MLE for $\sigma$. Details are in the Appendix~\ref{AppendixMLE}.

\subsubsection{Theoretical Coverage}\label{TheoCov}

Recall from Section~\ref{TCKH}, that the coverage under random
distortion is 
$$P\left(\frac{B^2}{A^2}\leqslant\frac{c_m}{c_k}\right)=P\bigg(c_m\left(x_2{}^T(H_{22}-H_{21}H_{11}{}^{-1}H_{12})x_2\right)\geqslant
c_k
\left(x_1{}^TH_{11}x_1+2x_1^TH_{12}x_2+x_2{}^T(H_{21}H_{11}{}^{-1}H_{12})x_2\right)\bigg)$$
where $x=(x_1,x_2)$ is the vector of random distortion and
$H=\left(\begin{array}{ll}H_{11} & H_{12}\\ H_{21}=H_{12}{}^T &
  H_{22}\end{array}\right)$ is the Fisher information matrix which is
assumed to be a constant, $m$ is the dimension of the parameters in
$\mathcal P$ and $m+k$ is the dimension of the larger model space
$\mathcal M$. Here the parameters in the equation above are $m=1$,
$k=J-1$. (Note that $x=(x_1,x_2)$ is $J+1$-dimensional since $x_1$ is
1-dimensional and $x_2$ is $J$-dimensional. However, the model is not
identifiable under these parameters, since rescaling the $p_i$
does not influence the resulting model. Therefore, $x_2$ only has $J$
degrees of freedom. The lost degree of freedom corresponds to an
eigenvector of this information matrix with eigenvalue 0.) If we let
$l$ be the expected log-likelihood with expectation taken under the
original normal distribution with mean 0, variance 1, then it is
straightforward to show (full details in Appendix~\ref{AppendixHessian}) that at $\tau=0$,
and $p_i=1$, $T_i=\frac{1}{J}$ we get:

\begin{align*}
\frac{\partial^2l}{\partial\sigma^2}&=-2\\
\frac{\partial^2l}{\partial\sigma\partial
  p_i}&=-\frac{1}{\sqrt{2\pi}}\left(c_{i-1}e^{-\frac{c_{i-1}{}^2}{2}}-c_{i}e^{-\frac{c_{i}{}^2}{2}}\right)\\
\frac{\partial^2l}{\partial p_i\partial
  p_j}&=\frac{1}{J^2}\qquad\textrm{if }i\ne j\\
\frac{\partial^2l}{\partial p_i{}^2}&=\frac{1}{J^2}-\frac{1}{J}\\
\end{align*}

Since in this simulation, the distortion is only in the $x_2$
direction, we have $x_1=0$ and, the adequate bootstrap interval should
cover the true value provided
\begin{align*}
c_{1}\left(x_2{}^T(H_{22}-H_{21}H_{11}{}^{-1}H_{12})x_2\right)&\geqslant
c_{J-1} \left(x_2{}^T(H_{21}H_{11}{}^{-1}H_{12})x_2\right)\\
\end{align*}
or equivalently
\begin{align*}
x_2{}^T\left(c_{1}(H_{22}-H_{21}H_{11}{}^{-1}H_{12})
-c_{J-1}(H_{21}H_{11}{}^{-1}H_{12})\right)x_2&\geqslant
0\\
\end{align*}
We see that $H_{22}=\frac{1}{J}I-\frac{1}{J^2}11^T$ and also
$H_{12}1=0$. Therefore, $1$ and $H_{21}$ are both eigenvectors of this
matrix, so picking an orthogonal basis containing $H_{21}$ and 1, the
matrix
$$c_{1}(H_{22}-H_{21}H_{11}{}^{-1}H_{12})
-c_{J-1}(H_{21}H_{11}{}^{-1}H_{12})$$
becomes diagonal with $J-2$ values $\frac{c_{1}}{J}$
(corresponding to vectors orthogonal to $1$ and $H_{21}$), one value 0
(corresponding to $1$),
and one value
$\frac{c_{1}}{J}-(c_{J-1}+c_1)H_{11}{}^{-1}H_{12}H_{21}$
(corresponding to $H_{21}$). 

So if the distortion is in a uniformly random direction (which is not
quite true in our log-normal simulation, but should be close enough
when the log-normal distribution has small $\tau$), the coverage
probability is the probability that
$(c_{J-1}+c_1)H_{11}{}^{-1}H_{12}H_{21}-\frac{c_{1}}{J}$ times a
chi-square distribution with one degree of freedom is less than
$\frac{c_{1}}{J}$ times a chi-square distribution with $J-2$
degrees of freedom.  Let $X$ follow a chi-squared distribution with 1
degree of freedom and $Y$ follow a chi-squared distribution with
$J-2$ degrees of freedom. Then we have

\begin{align*}
P\left(\left((c_{J-1}+c_1)H_{11}{}^{-1}H_{12}H_{21}-\frac{c_{1}}{J}\right)X\leqslant
\frac{c_{1}}{J}Y\right)&=P\left(\frac{Y}{(J-2)X}\geqslant
\frac{\left((c_{J-1}+c_1)H_{11}{}^{-1}H_{12}H_{21}-\frac{c_{1}}{J}\right)}{(J-2)\left(\frac{c_{1}}{J}\right)}\right)\\
&=P\left(\frac{Y}{(J-2)X}\geqslant
\frac{\left(J(c_{J-1}+c_1)H_{11}{}^{-1}H_{12}H_{21}-c_{1}\right)}{(J-2)c_{1}}\right)\\
\end{align*}

That is, the probability that an F distribution with parameters $J-2$
and 1 is greater than
$\frac{\left(J(c_{J-1}+c_1)H_{11}{}^{-1}H_{12}H_{21}-c_{1}\right)}{(J-2)c_{1}}$. We
calculate these coverage probabilities for model adequacy tests with
size 5\%, for the values of $J$ used in our simulation. The coverage
probabilites (multiplied by 1000) are given in
Table~\ref{SampBiasSimResCover} for comparison with the simulation
results.

It is worth noting that in practical situations, we would not have
such detailed knowledge of the distribution of the true sampling
bias. If we did, we would be able to fit a much better model to the
data, and potentially use some Bayesian methods to achieve better
results than our adequate bootstrap. The main purpose of the adequate
bootstrap is that it is robust against whatever sources of model
uncertainty may exist. In the cases where we do not know the sources
of model uncertainty, we cannot use the above methods to estimate the
coverage probabilities.

We also have from Section~\ref{TCKH} that the adequate bootstrap size
is
$\frac{c_{k}}{x_2{}^T(H_{22}-H_{21}H_{11}{}^{-1}H_{12})x_2}$. Diagonalising
$H_{22}-H_{21}H_{11}{}^{-1}H_{12}$, we have that the eigenvalues are $J-2$
eigenvalues of $\frac{1}{J}$, one eigenvalue of 0 and one eigenvalue
of $\frac{1}{J}-H_{12}H_{21}H_{11}{}^{-1}$. The adequate bootstrap size
should therefore be $\frac{c_{J-1}}{X}$, where $X$ is the sum of
$\frac{1}{J}-H_2H_2{}^TH_1{}^{-1}$ times a chi-squared distribution
with one degree of freedom plus $\frac{1}{J}$ times a chi-squared
distribution with $J-2$ degrees of freedom. For $J=5$, this is
0.1063486465 times a chi-square with one degree of freedom, plus 0.2
times a chi-square distribution with 3 degrees of freedom.

%
%
%
%
%

\subsection{Results}

\begin{table}[htbp]
\caption{Coverage of adequate bootstrap confidence interval and
  standard bootstrap confidence intervals for
  various settings of number of intervals $J$ and values of
  $\tau$. Results are out of 1000
  simulations.}\label{SampBiasSimResCover}

\hfil\begin{subtable}{0.4\textwidth}
\caption{Adequate Bootstrap}
\begin{tabular}{r|rrrr}
\hline
\backslashbox{$J$}{$\tau$}    &  0 &  0.2 & 0.5 & Theory\\
\hline
  3 &     & 998 & 989 & 1000\\
  5 & 953 & 953 & 927 & 915\\
  8 &     & 943 & 937 & 868\\
\hline
\end{tabular}
\end{subtable}\hfil\begin{subtable}{0.28\textwidth}\caption{Standard
  Bootstrap}
\begin{tabular}{r|rrrrr}
\hline
\backslashbox{$J$}{$\tau$}   &   0 & 0.2 & 0.5 \\
\hline
  3 &     & 196 & 71 \\
  5 & 952 & 164 & 55 \\
  8 &     & 203 & 70 \\
\hline
\end{tabular}
\end{subtable}
\end{table}

Table~\ref{SampBiasSimResCover} compares the coverage of the adequate
bootstrap confidence interval out of 1000 simulations in each
scenario, with the adequate bootstrap size calculated
nonparametrically with the size of the model adequacy test fixed at
0.05, in comparison to the standard bootstrap confidence interval for
this simulation.  We see that the coverage for a standard bootstrap is
fairly poor. Also note that in the null model case, the adequate
bootstrap size is the full 20,000 data points, and the coverage
matches the standard bootstrap.  We see the observed coverage is a
little better than our calculated coverage from the previous
section. There are a few sources of error in the calculations from the
previous section. Firstly, the calculated coverage was based on a
quadratic approximation to the likelihood, which is not
perfect. Secondly, the calculated coverage was based on the direction
of distortion being uniformly distributed on the sphere, which is the
case for normal distortion, but only approximate for the log-normal
distortion we used here. Finally, the theoretical coverage is based on
a theoretical calculation of the adequate bootstrap
size. Table~\ref{SampBiasSimResABS} gives the mean and median
bootstrap sizes for the simulation scenarios. The mean is much larger
than the median, representing the fact that in a small number of
simulations where the simulation distribution is closer to the true
distribution, the adequate bootstrap size can be very large. (We limit
it to the size of the data set, since bootstraps of this size produce
the amount of uncertainty present in the data, so even if there is no
uncertainty due to model misspecification, we cannot obtain a better
confidence interval.)

\begin{table}
\caption{mean and median adequate bootstrap sizes for various settings of number
  of classes and values of $\tau$. Results are out of 1000
  simulations.}\label{SampBiasSimResABS}

\hfil\begin{subtable}{0.3\textwidth}\caption{Mean}
\begin{tabular}{c|rrr}
\hline
\backslashbox{$J$}{$\tau$}   &  0 & 0.2 & 0.5 \\
\hline
  3 &           & 1024 & 234 \\
  5 & 19200 & 424 & 73 \\
  8 &           & 335 & 57 \\
\hline
\end{tabular}
\end{subtable}\hfil\begin{subtable}{0.3\textwidth}\caption{Median}\begin{tabular}{c|rrr}
\hline
\backslashbox{$J$}{$\tau$}   &  0 & 0.2 & 0.5 \\
\hline
  3 &       &  480 & 84 \\
  5 & 20000 &  359 & 61 \\
  8 &       &  328 & 54 \\
\hline
\end{tabular}
\end{subtable}\end{table}

\begin{figure}[!htbp]

\begin{subfigure}{0.5\textwidth}

\hfil\includegraphics[height=8cm,clip=true,trim=0cm 0cm 1cm 1cm]{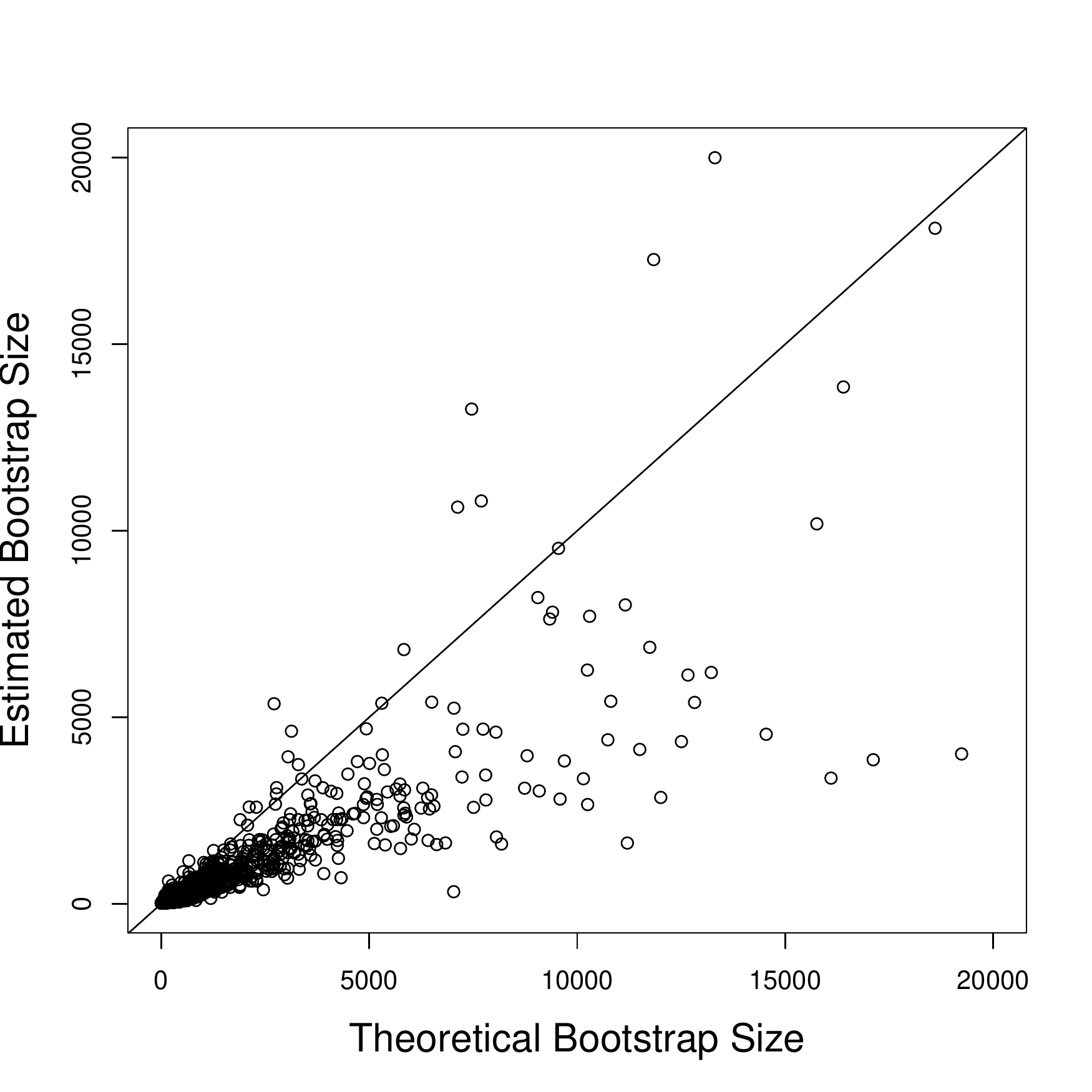}
\caption{Sample Size 20,000}
\end{subfigure}
\begin{subfigure}{0.5\textwidth}
\hfil\includegraphics[height=8cm,clip=true,trim=0cm 0cm 1cm 1cm]{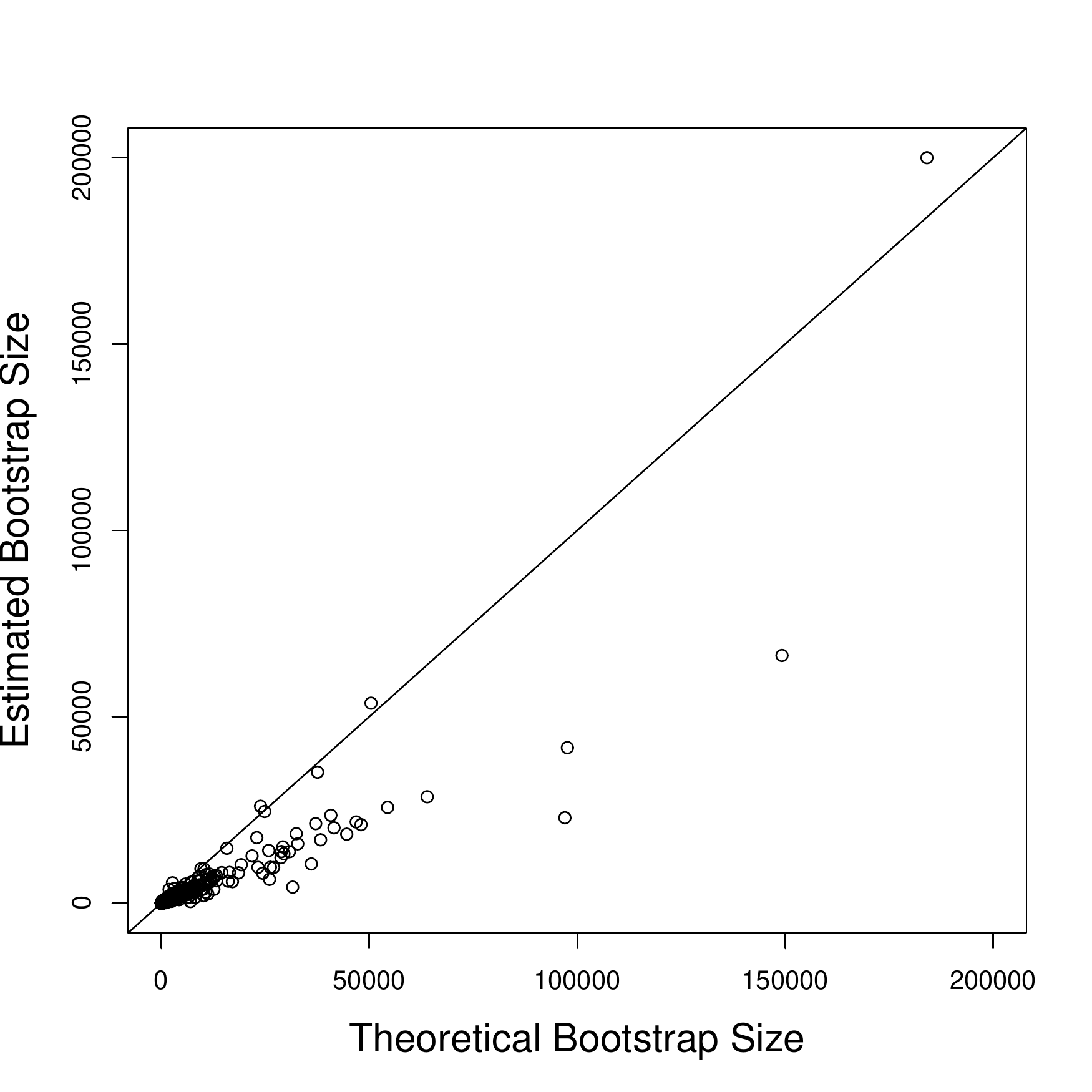}
\caption{Sample Size 200,000}\label{BootstrapsizesLong}
\end{subfigure}

\caption{Theoretical versus empirically estimated adequate bootstrap
  sizes for the simulations with $J=3$ and $\tau=0.2$ or
  $\tau=0.5$}\label{Bootstrapsizes}

\end{figure}

The finite sample size, and the empirical (and slightly heuristic)
method used to estimate the adequate bootstrap size mean that the
actual bootstrap size used may differ substantially from the
theoretical size. Our calculations in the previous section give us a
theoretical value of the adequate bootstrap size for each simulated
set of sampling bias probabilities $p_i$. Figure~\ref{Bootstrapsizes}
compares this theoretical adequate bootstrap size with the empirically
estimated size for the simulations with $J=3$ and $\tau=0.2$ or
$\tau=0.5$ (a total of 2000 simulations).

As explained in Section~\ref{FiniteSample}, it was pointed out by
\citet{Lindsay} for a bootstrap with replacement from a
finite sample, the probability of rejecting the null model is slightly
higher than for a sample from the underlying distribution, so the
adequate bootstrap size should be slightly lower, on average, than the
theoretical value. This is consistent with
Figure~\ref{Bootstrapsizes}, and also explains why the coverage is
better than expected in the $J=5$ and $J=8$ cases. The theory also
predicts that the difference between theoretical and estimated
adequate bootstrap size should get larger as the theoretical bootstrap
size approaches the size of the data set. This effect is also seen in
Figure~\ref{Bootstrapsizes}. As the sample size
becomes larger, the empirical sample should become closer to the true
distribution, and the estimated bootstrap size should be closer to the
theoretical bootstrap size. We confirm this by performing a further
simulation in which the size of the data set is increased to
200,000. This should result in more accurate estimated bootstrap
sizes, and therefore coverage closer to the theoretical coverage. 

%
%
%
%

%
%

\begin{figure}[!htbp]
\hfil\includegraphics[height=7cm,clip=true,trim=0cm 0cm 1cm 1cm]{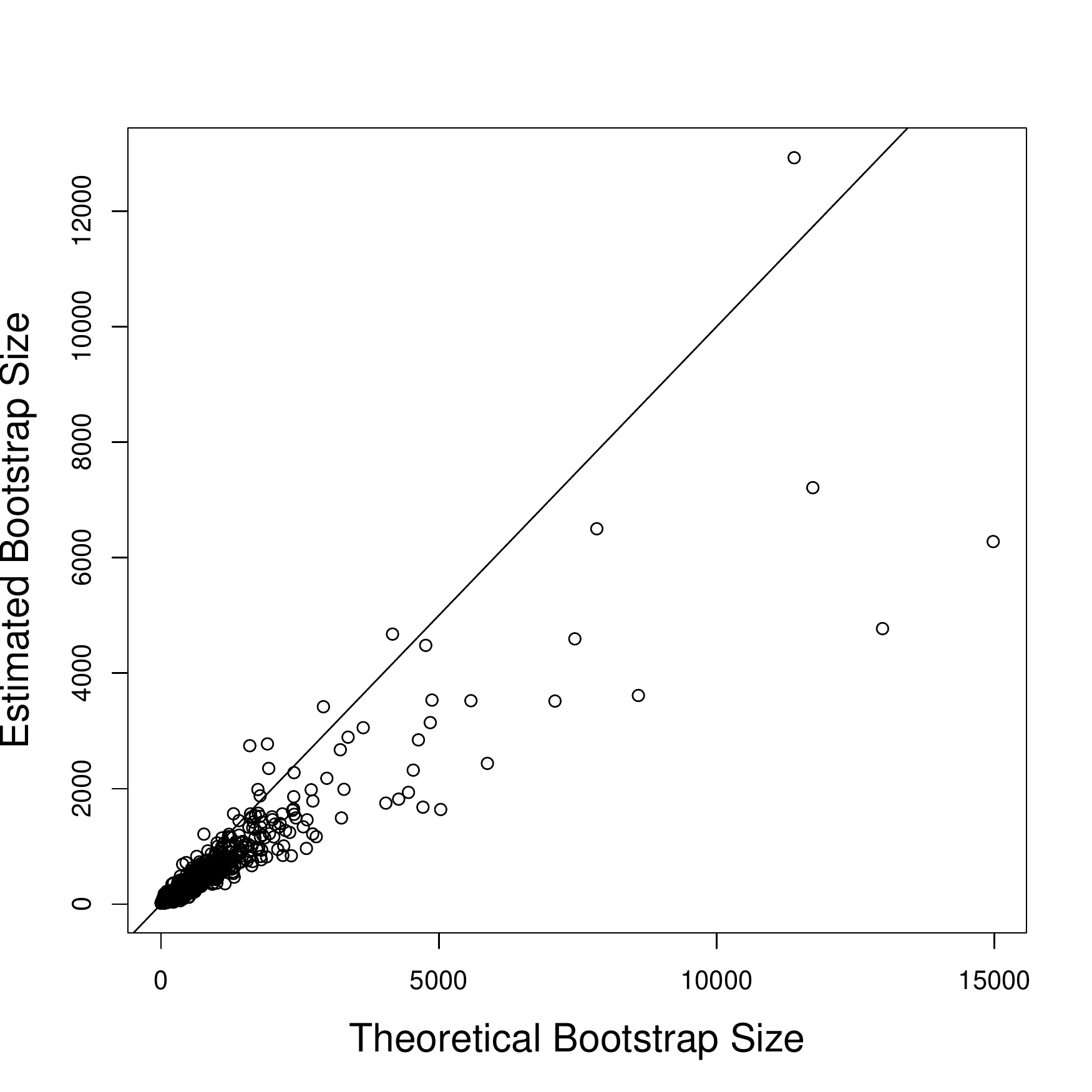}
\caption{Theoretical versus empirically estimated adequate bootstrap
  sizes for the simulations with $J=5$ and $\tau=0.2$ or
  $\tau=0.5$ for sample size 20,000}\label{BootstrapsizesLong5}

\end{figure}

\begin{figure}
\hfil\begin{subfigure}{0.37\textwidth}\caption{$p_i=1$}
  \includegraphics[width=5cm,clip=T,trim=1cm 1cm 2cm
    2cm]{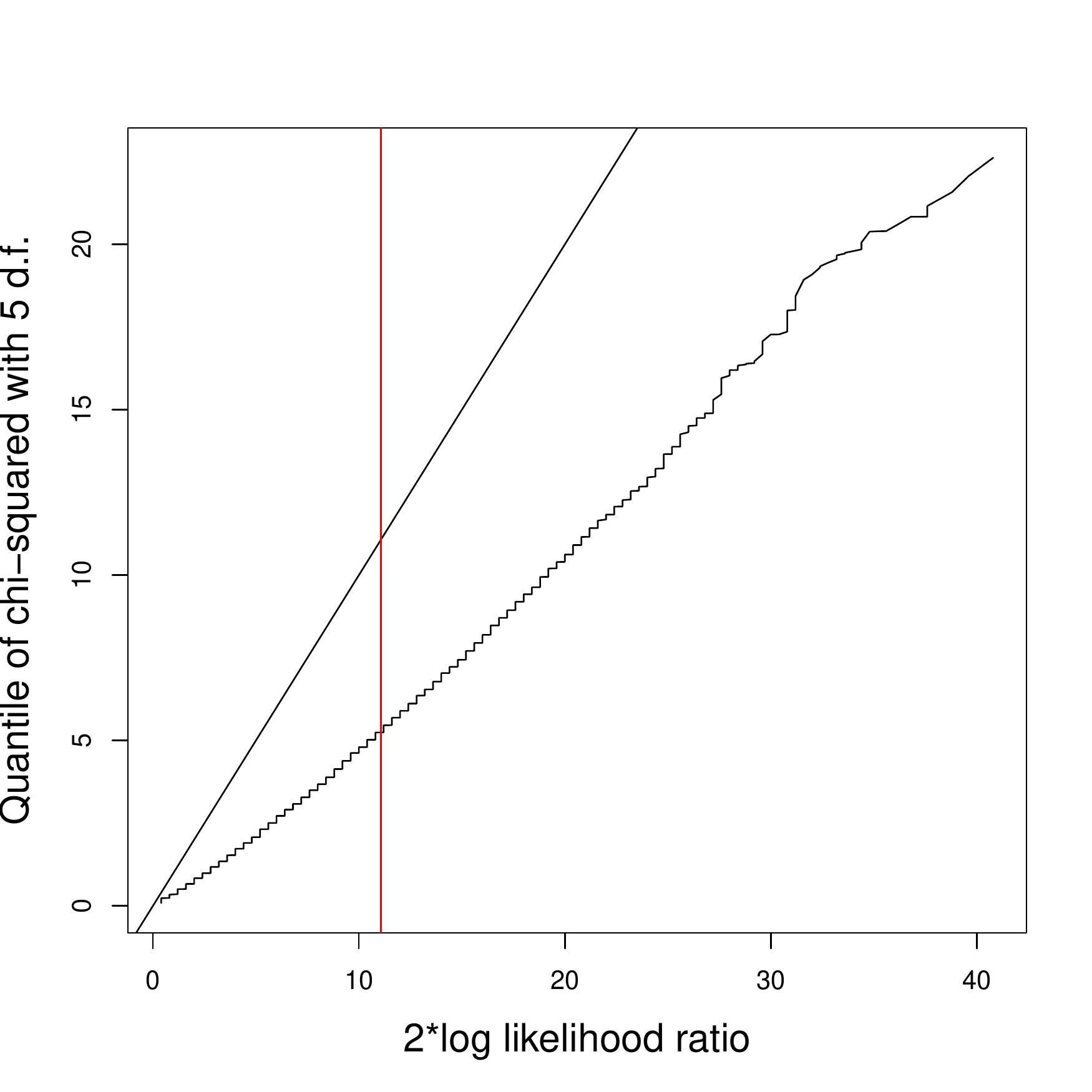}
\end{subfigure}\hfil%
\begin{subfigure}{0.37\textwidth}\caption{$p=(1,0.96,1.01,0.95,1)$}
  \includegraphics[width=5cm,clip=T,trim=1cm 1cm 2cm
    2cm]{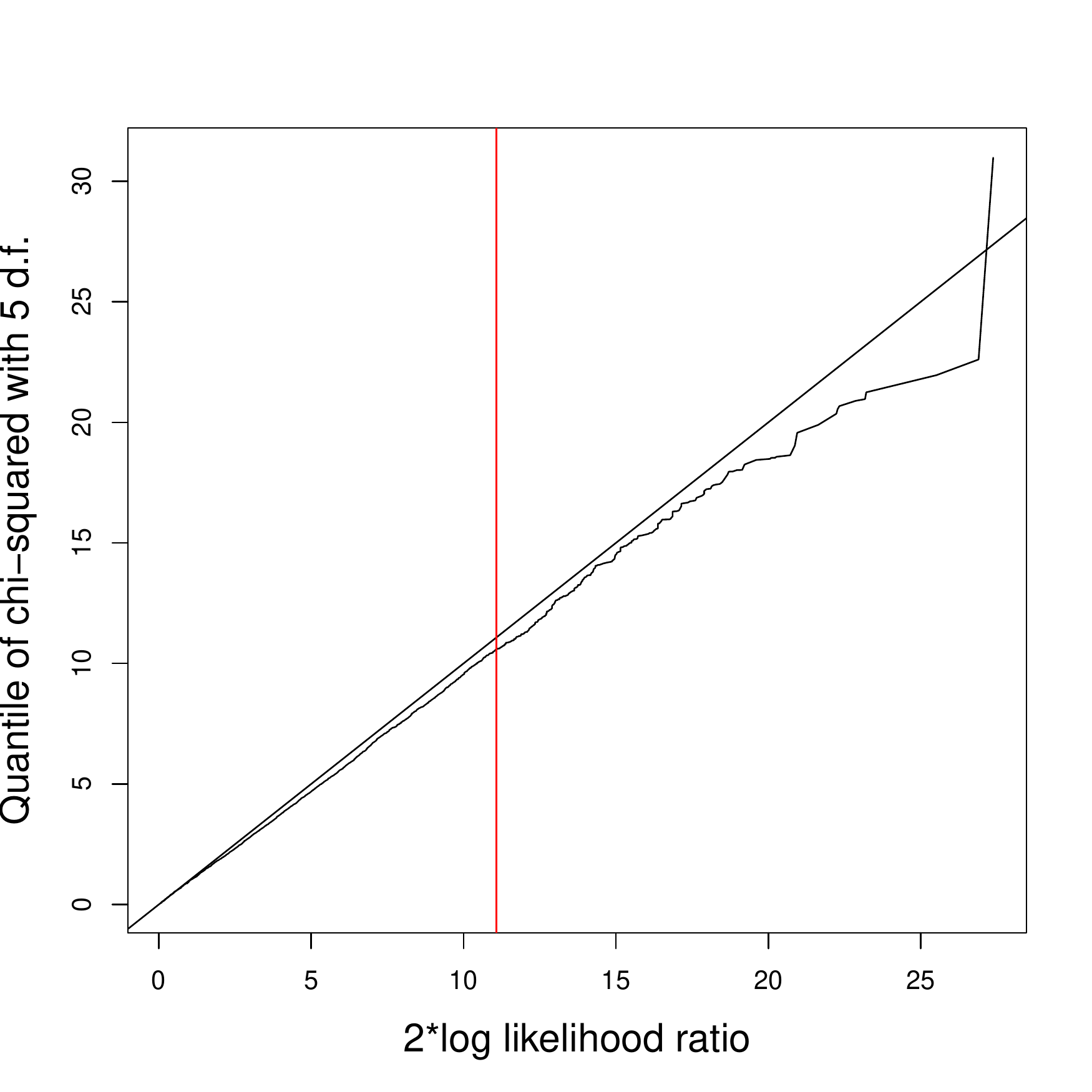}
\end{subfigure}%
%

\caption{$Q$--$Q$ plots of log-likelihood ratio statistic
  between the true distribution parameters and the MLE
  parameters. Data from sample of size 200000; bootstraps of size
  200000 for the null case and 9600 for the non-null case (the
  calculated adequate bootstrap sizes for these cases). Red lines
  indicate the 5\% critical value of the theoretical chi-squared
  distribution.}\label{QchiSqLLRat} 

\end{figure}

On the other hand, for the $J=5$ case there is a bigger disparity
between the estimated and theoretical adequate bootstrap sizes, as
shown in Figure~\ref{BootstrapsizesLong5}. Here we see that there is a
clear bias towards underestimating the bootstrap sizes, resulting in a
better coverage than expected. The problem with our theory is that it
is based on the asymptotics of log-likelihood ratio
statistics. However, these assymptotics do not hold so closely for
smaller samples. It turns out that the finite-sample MLE for $\sigma$
is biassed, so if the adequate bootstrap size is small, then our
bootstrap estimates for $\sigma$ could be biassed, even in the case
without sampling bias, making the assymptotic formula for the
likelihood inaccurate for this finite sample, and making our estimates
of the coverage inaccurate. This is demonstrated in
Figure~\ref{QchiSqLLRat}, which compares the distribution of the
log-likelihood ratio statistic with the theoretical chi-squared
distribution that it is expected to follow. We see that this statistic
is substantially higher than the theoretical chi-squared statistic it
is assumed to follow. This results in the likelihood ratio test having
higher power than our theory predicts (and higher size than the value
we set), and so the theoretically predicted adequate bootstrap size is
too large, and as a consequence, the theoretical coverage is too
small. This matches the observed results. It is worth noting here that
often the differences between asymptotic estimates and finite-sample
estimates can be remedied by taking a larger sample. However, for the
adequate bootstrap, it is the bootstrap subsamples that would need to
be made larger, but their size is determined by the goodness-of-fit
between the data and the model. Therefore, we are stuck with the
finite sample size, and cannot always make use of asymptotic
approximations by collecting more data.

Proponents of subsampling instead of bootstrapping might argue that
the reason for the bias could be the use of a bootstrap in place of a
subsample, rather than the finite sample asymptotics.  Indeed, we see
that the asymptotic distribution fits the observed distribution far
less well in the bootstrap sample on the left, which can be explained
by the fact that a bootstrap sample is not an independent sample from
the underlying distribution, with repeated observations making a
bigger distortion in the likelihood. However, even in the case where
our boostrap contains only 9,600 samples out of 200000, we see a
noticeable difference between the distributions. In this case, it is
unlikely that a bootstrap sample includes any repeated observations,
so the bootstrap will approximate a sample from the underlying
distribution very well in this case. However, we still see that the
observed distribution of the log-likelihood ratio statistic does not
match the theoretical chi-squared distribution.  

We demonstrate that using subsampling instead of bootstrapping does
not fix the problem of coverage not matching the theoretical values
(at least in this case) 
%
by rerunning the adequate bootstrap using subsampling without
replacement in place of bootstrap with replacement. We found that this
change does not produce significantly different coverages for these scenarios
(Table~\ref{SampBiasSimResCoverJackknife} in Appendix~\ref{AppendixTables}).

In addition to coverage, we look at the width of the confidence
intervals in Table~\ref{SampBiasSimResWidth}. We see that in the null
case, the width is very similar to the width of an ordinary bootstrap
interval. As the distribution gets further from the model
distribution, the adequate bootstrap size decreases, and the width of
the confidence interval increases, reflecting the greater uncertainty
in the underlying parameter values.

\begin{table}[htbp]
\caption{Mean confidence interval widths for
  various settings of number of intervals $J$ and values of
  $\tau$. Results are out of 1000 simulations.}\label{SampBiasSimResWidth}

\begin{subtable}{0.4\textwidth}
\caption{Adequate Bootstrap}
\hfil\begin{tabular}{r|rrr}
\hline
\backslashbox{$J$}{$\tau$}    &  0 & 0.2 & 0.5 \\
\hline
  3 &       & 0.341 & 0.720 \\
  5 & 0.040 & 0.351 & 0.775 \\
  8 &       & 0.343 & 0.799 \\
\hline
\end{tabular}
\end{subtable}\hfil
\begin{subtable}{0.4\textwidth}
\caption{Standard Bootstrap}
\begin{tabular}{r|rrr}
\hline
\backslashbox{$J$}{$\tau$}  &   0 & 0.2 & 0.5 \\
\hline
  3 &        & 0.0391 & 0.0388 \\
  5 & 0.0392 & 0.0391 & 0.0386 \\
  8 &        & 0.0391 & 0.0385 \\
\hline
\end{tabular}
\end{subtable}
\end{table}

The conclusion from this simulation is that the adequate bootstrap
does provide a confidence interval that incorporates the parameter
uncertainty due to model misspecification; however, even in the most
ideal circumstances, it is difficult to control the coverage of the
confidence interval in the way that we are able to do under the
assumption that the model is the true model.

\section{Real Data Analysis}\label{REALDATA}

\subsection{European Stock Markets}

We demonstrate the application of the adequate bootstrap method on three
real data sets. The first is the European stock markets data set. The
closing prices of four major European stock indices during the period
1991--1998 are available in the {\tt R} dataset {\tt EuStockMarkets}
(provided by Erste Bank AG, Vienna, Austria). From this dataset, we
extracted daily and weekly gains by taking the ratio of each day's
closing price to the closing price one (respectively five) days
earlier. These gains were not corrected for holidays, so do not always
correspond perfectly to calendar weeks. For weekly gains, we used
non-overlapping weeks, so we obtained 371 weekly gains over the 1860
business day period.

A common assumption in finance is that stock market gains follow a
log-normal distribution. Under this assumption, we used an adequate
bootstrap to obtain a 95\% confidence interval for the Value at Risk
of the distribution at the 99\% level. Value at Risk is the term in
finance for the percentiles of the distribution, so the Value at Risk
at the 99\% level means the 1st percentile of the distribution. (For
this example we usually consider our risks to be losses, so the 1st
percentile of the returns distribution is the 99th percentile of the
loss distribution.) Value at Risk is an important measure in risk
assessment, and is commonly used by financial firms to measure the
risk of a position, and to set appropriate capital reserves.

It is of course straightforward to estimate percentiles of a
distribution non-parametrically. Usually, the benefits of the adequate
bootstrap are clearer when the quantity to be estimated does not have
a non-parametric meaning. However, since the exact value of the
percentile depends on the lower tail of the distribution,
non-parametric estimators can be very inefficient compared with
parametric ones. Therefore, if the log-normal distribution fits the
data well, then we expect to obtain a tighter confidence interval for
this Value at Risk than we can obtain with the nonparametric
estimator.

\begin{figure}[htbp]

\begin{subfigure}{0.5\textwidth}
\includegraphics[width=8cm]{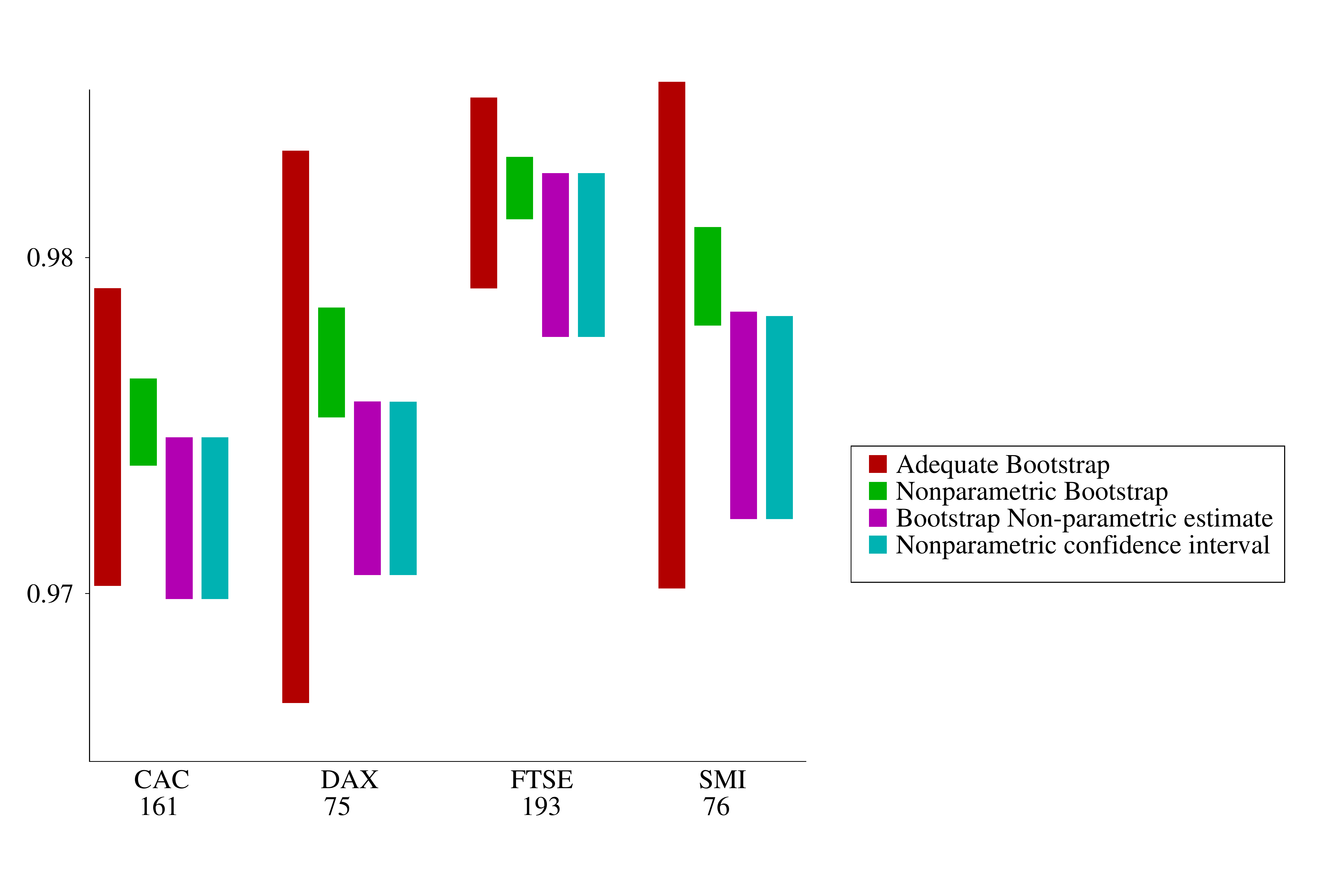}
\caption{Daily Returns}
\end{subfigure}
\begin{subfigure}{0.5\textwidth}
\includegraphics[width=8cm]{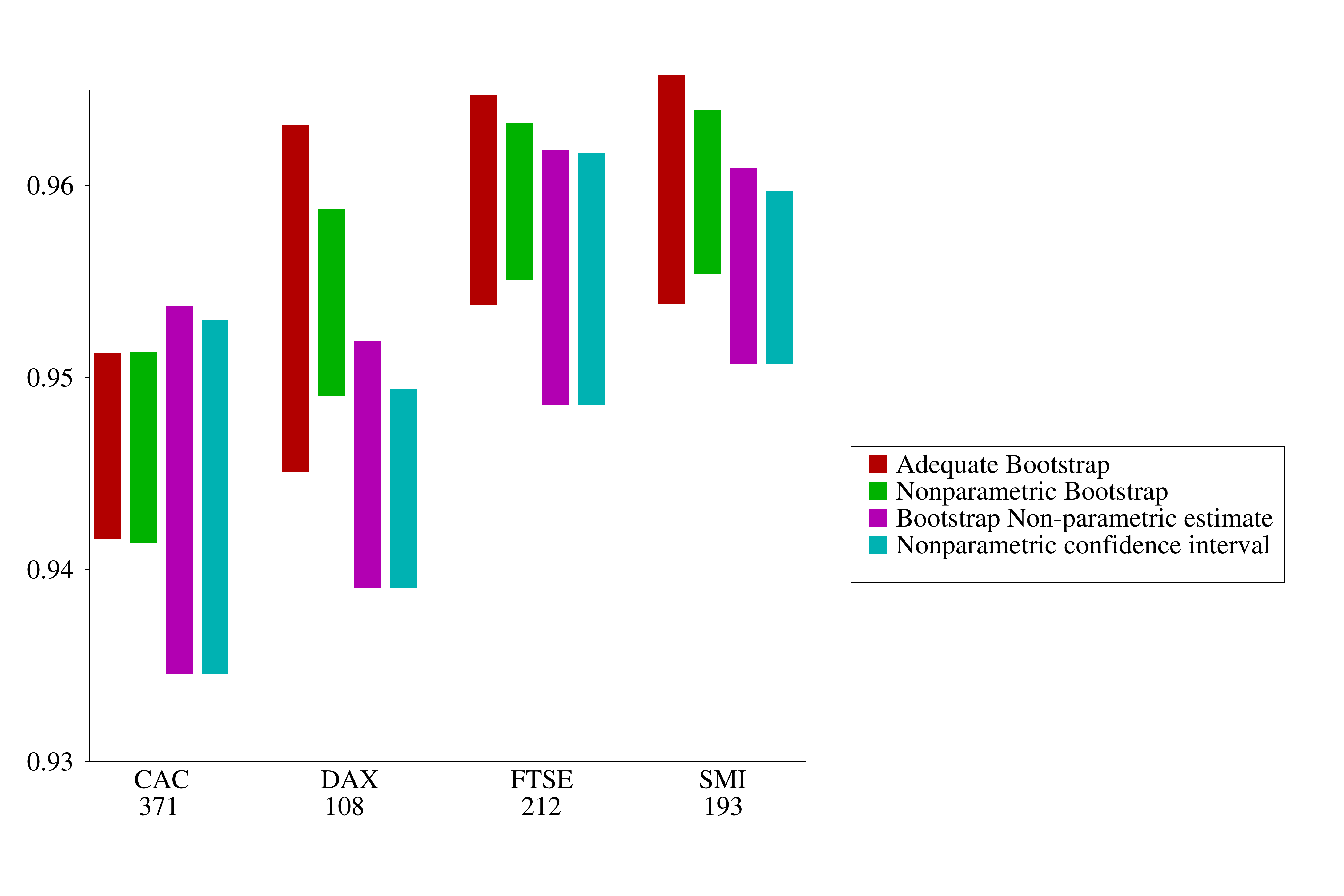}
\caption{Weekly Returns}
\end{subfigure}

\caption{Confidence intervals for 99\% Value at Risk of European Stock
  Indices in 1991--1998 period for different estimating
  methods. Numbers under the name of each index give the estimated
  adequate bootstrap sizes.}\label{EuStock}

\end{figure}

We compare the results of adequate bootstrap with a non-parametric
bootstrap for estimates of the log-normal parameters. Since the Value
at Risk is a distributional quantity, not depending on any particular
parametric distribution, we also have a non-parametric estimator
formed by taking the first percentile of the observed data. We used a
non-parametric bootstrap to obtain a confidence interval for this
non-parametric estimator. Finally, there is a non-parametric
confidence interval for the Value at Risk, obtained by observing that
the number of observations in a sample of size $n$ that lie below the
1st percentile of the distribution is binomial with parameters $n$ and
$0.01$, so a 95\% confidence interval for this number can be found as
$[n_l,n_u]$. Now the $n_l$th order statistic and the $(n_u+1)$th order
statistic form a 95\% confidence interval for the Value at Risk.  The
confidence intervals estimated by these methods are shown in
Figure~\ref{EuStock}.

We see that for the daily returns, the adequate bootstrap size is
fairly small, and the each adequate bootstrap confidence interval is
larger than the corresponding non-parametric interval. This indicates
that the log-normal is not a good fit for the data, and that the
non-parametric confidence intervals are probably more reliable. The
only case where the confidence intervals are close in size is the
FTSE. In this case the confidence interval based on the adequate
bootstrap log-normal seems reasonable. In all cases, the interval
based on a standard non-parametric bootstrap and the log-normal model
is far too small, as the log-normal model does not fit the data well,
and performing a bootstrap on a sample size that is too large for the
fit of the data to the model leads to falsely reassuring confidence
intervals.

For the weekly results, the log-normal model is slightly more
reasonable. Indeed for the CAC, the log-normal model is found
adequate. Also, with such a small data set, the non-parametric
interval is very wide. Therefore, the adequate bootstrap is
reasonable, and finds a narrower confidence interval, since 
non-parametric estimators of Value at Risk are inefficient. For the
DAX, the adequate bootstrap size is small, so the confidence interval
is wide, and the non-parametric interval is more appropriate. For the
FTSE and the SMI, the adequate bootstrap size is moderate, and the
adequate bootstrap interval and the nonparametric interval are of
comparable length. The adequate bootstrap interval is slightly higher,
which could be because the observed distribution has a heavier tail
than the log-normal. It could also be because the non-parametric
interval is asymmetric. The choice of confidence interval in these
cases would depend largely on how much confidence we place on our
log-normal assumption to start with. If we are very confident that a
large proportion of the gains are driven by some mechanism that should
lead to a log-normal distribution, then we should prefer the adequate
bootstrap interval. If the log-normal distribution was a wild guess,
then the non-parametric confidence interval might be more appealing.

\subsection{Power Law Distributions}

For a second real data analysis, we look into power law
distributions. There have been a number of distributions claimed to
follow power law distributions across a range of
subjects. Furthermore, a number of natural models of dynamical systems
based on a ``rich get richer'' type of effect, naturally lead to power
law type distributions. This makes power law distributions a
relatively appealing candidate for adequate bootstrap analysis for the
following reasons:

\begin{itemize}

\item There are a number of data sets where the power law has been
  suggested as appropriate, with suggested dynamic mechanisms that
  lead to a power law. 

\item The power parameter in the power law does not directly
  correspond to a distributional quantity that can be estimated
  non-parametrically. 

\item However, the power parameter sometimes corresponds directly to a
  meaningful parameter in the hypothesized generating process, so a
  confidence interval for this parameter is a meaningful inference.

\item The power parameter is also linked to existing measures of
  inequality, such as the Gini coefficient (particularly in the
  context of wealth distribution).

\end{itemize}

\begin{figure}

\hfil\hfil\includegraphics[width=10cm]{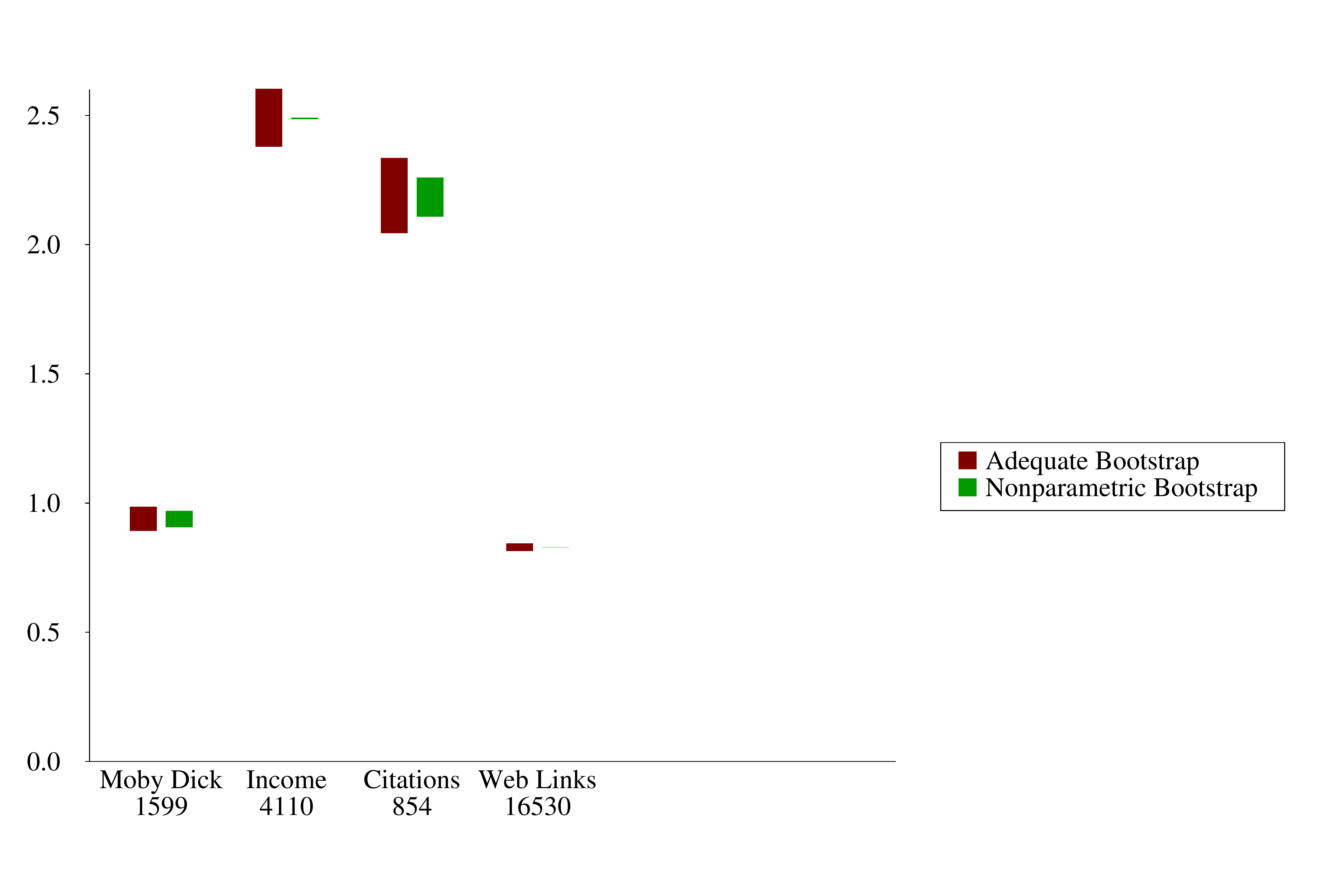}
\caption{Adequate bootstrap and nonparametric bootstrap 95\%
  confidence intervals for the power in a power law distribution for
  various datasets. Numbers below the dataset name indicate the
  adequate bootstrap size for this power law and this data set.}\label{PowerLawFig}
\end{figure}

We look at a number of datasets that have been claimed to follow a
power law in Figure~\ref{PowerLawFig}. The datasets considered in this
analysis are summarised in Table~\ref{PowerLawDatasets}. Full details
of the sources of these data sets are in
Table~\ref{PowerLawDatasetsSources} in Appendix~\ref{AppendixTables}

\begin{table}[htbp]

\caption{Data sets used for power law analysis.}\label{PowerLawDatasets}

\begin{tabular}{ll}
\hline
Name& Description \\
\hline
SNP& Market Capitalization of SNP500 companies.\\
City & population of cities in England and Wales.\\
Country & populations of 34 countries in 2011.\\
Moby Dick & Frequencies of words in the novel {\em Moby Dick}. \\
Citations&Number of ISI citations for a paper (1981--1997). \\
Web Links&Number of Links to a website (1999).\\
Quakes&Magnitude of earthquakes on the Richter scale.\\
Income&Canadian individual income in 2013 for individuals\\
\hline
\end{tabular}

\end{table}

Some of the
datasets were truncated to focus on the tail following a power
law. Three of the datasets were treated as individual data following a
type-1 Pareto distribution. Given the size of the observations, this
seemed reasonable even though some observations were technically counts. 
The income data was grouped. Four other data sets (Moby Dick, Citations, Web
links, and Earthquakes) needed to be treated as discrete --- the first
three are count data including many small counts; the fourth is rounded
to one decimal place. We treated all three as rounded data, and
converted to grouped data, mostly with intervals $[n-0.5,n+0.5]$, but
some of the larger values were grouped in wider intervals, so that a
chi-squared test can be more reasonably carried out.

\begin{table}[htbp]

\caption{Results for power law analysis.}\label{PowerLawResults}

\setlength{\tabcolsep}{0.1cm}
\hfil\begin{tabular}{llrrrcc}
\hline
&Dataset& Lower limit & Data set  & Adequate & Adequate Bootstrap & Standard Bootstrap\\
&&&size & Bootstrap size & confidence interval & confidence interval\\
\hline
\multirow{3}{5em}{\parbox{5em}{Individual data}}&SNP       &1750.0&504& 20   &  $[0.343, 0.493]$ & $[0.391, 0.421]$ \\
&City      &5000.0&416& 69   &  $[0.554, 0.804]$ & $[0.613, 0.713]$ \\
&Country   &300000.0&34& 20   &  $[0.223, 0.316]$ & $[0.231, 0.302]$ \\
\hline
\multirow{5}{5em}{\parbox{5em}{Grouped data}} 
&Income    &50000.0&7984940& 4110 &  $[2.378, 2.603]$ & $[2.486, 2.491]$ \\
&Moby Dick &6.0& 3427& 1599 & $[0.892, 0.986]$ & $[0.906, 0.970]$ \\
&Citations &166.0&3210& 854 & $[2.044, 2.336]$ & $[2.108, 2.260]$\\
&Web Links &1.0&241428853& 16530 &  $[0.814, 0.844]$ & $[0.829, 0.829]$ \\
&Quakes    &3.5&5910& 298  &  $[6.409, 7.819]$ & $[6.905, 7.218]$ \\
\hline
\end{tabular}

\end{table}

The results are shown in Table~\ref{PowerLawResults}. We see that for
some data sets, such as the SNP and the Country datasets, the power
law is totally inappropriate. For the city populations and Moby Dick
word frequency datasets, the power law can be rejected by a relatively
small sample size. This makes its use questionable for these data
sets. The adequate bootstrap interval is fairly wide, indicating that
we cannot get a sufficiently accurate parameter estimate for this
distribution. For the grouped data, we see that even though adequate
bootstrap size is much larger, the confidence intervals are still
wide. This is because observations of grouped data are much less
informative, so while we need more observations to reject the model,
this does not indicate that the model is necessarily a better
fit. Judging from the confidence interval, it seems that the relative
error for the income data set is moderate, so the model may be
acceptable for these data, though the goodness-of-fit is limited. For
the Earthquake data, the adequate bootstrap size is nearly 300, but
the confidence interval is wide. This indicates that the model is not
appropriate. For the web links data, the adequate bootstrap size is
large, and the confidence interval is relatively small. This indicates
that the power law distribution is likely to be a useful model in this
case. Some of these confidence intervals are shown in
Figure~\ref{PowerLawFig}. 

The last four of these examples were studied in
\citet{PowerLaw}. Their results confirm that the
earthquake data does not fit the hypothesised power law. They also
reject the power law for the web links data, but this is more a
consequence of the size of the data set than the model not being
suitable for the data. Interestingly, they truncate the data at 3684,
and estimate the power parameter $\alpha$ as 1.336, which is outside
our confidence interval. (For the Pareto distribution we used here,
$\alpha$ is the exponent of $x$ in the survival function. In
\citet{PowerLaw}, they took the power parameter as the
exponent of $x$ in the density function, so their estimates differ
from ours by 1. When quoting their results, we have subtracted 1 to
convert to our choice of parameter, so for example for the web links
data set, they described the exponent as 2.336, and we converted to
the corresponding $\alpha$ parameter of 1.336.)  This indicates that
while the power law fits the data rather well, it does not fit the
tail so well, and that around the tail, a different exponent is
appropriate for the power law. Observe that the tail of the data that
they studied included only 28,986 out of 241,428,853 websites, so a
typical adequate bootstrap of size 16530, would contain an average of
$\frac{16530\times 28986}{241428853}=1.98$ data points in this tail,
so the fact that this small tail acts like a power law with a
different exponent does not prevent the power law we identified from
being a good model for the majority of websites. As we see, a large
number of data points are needed to falsify the model. They also find
the power law a good fit for the word frequencies in Moby Dick. We
reject the power law as not being a perfect fit, but still find a
large adequate bootstrap size, and a reasonably short confidence
interval. We therefore agree that the power law distribution should
prove useful for that data. Finally for Citations, again, we are able
to reject the power law model, but agree that it is a moderate
fit. From the adequate bootstrap confidence interval, we see that
there is still substantial uncertainty about the parameter estimate,
which may limit the usefulness of the model.

These examples highlight the advantages of the adequate bootstrap over
the credibility index of \cite{Lindsay}. For these data sets,
the credibility index of around 300 for the earthquake data might
suggest that the model is useful for this dataset. Only with the
adequate bootstrap highlighting the corresponding uncertainty due to
model misspecification do we see that the model is not such a good
fit here. For the income data, a credibility index over 4,000 might
suggest that the model is an extremely good fit for this
dataset. However, when we look at the adequate bootstrap confidence
interval, we see that there is still substantial uncertainty about the
relevant parameter.

\subsection{Income in European Countries}

\begin{figure}[htbp]
\hfil\includegraphics[width=9cm,clip=true,trim=0cm 0cm 0cm 1.5cm]{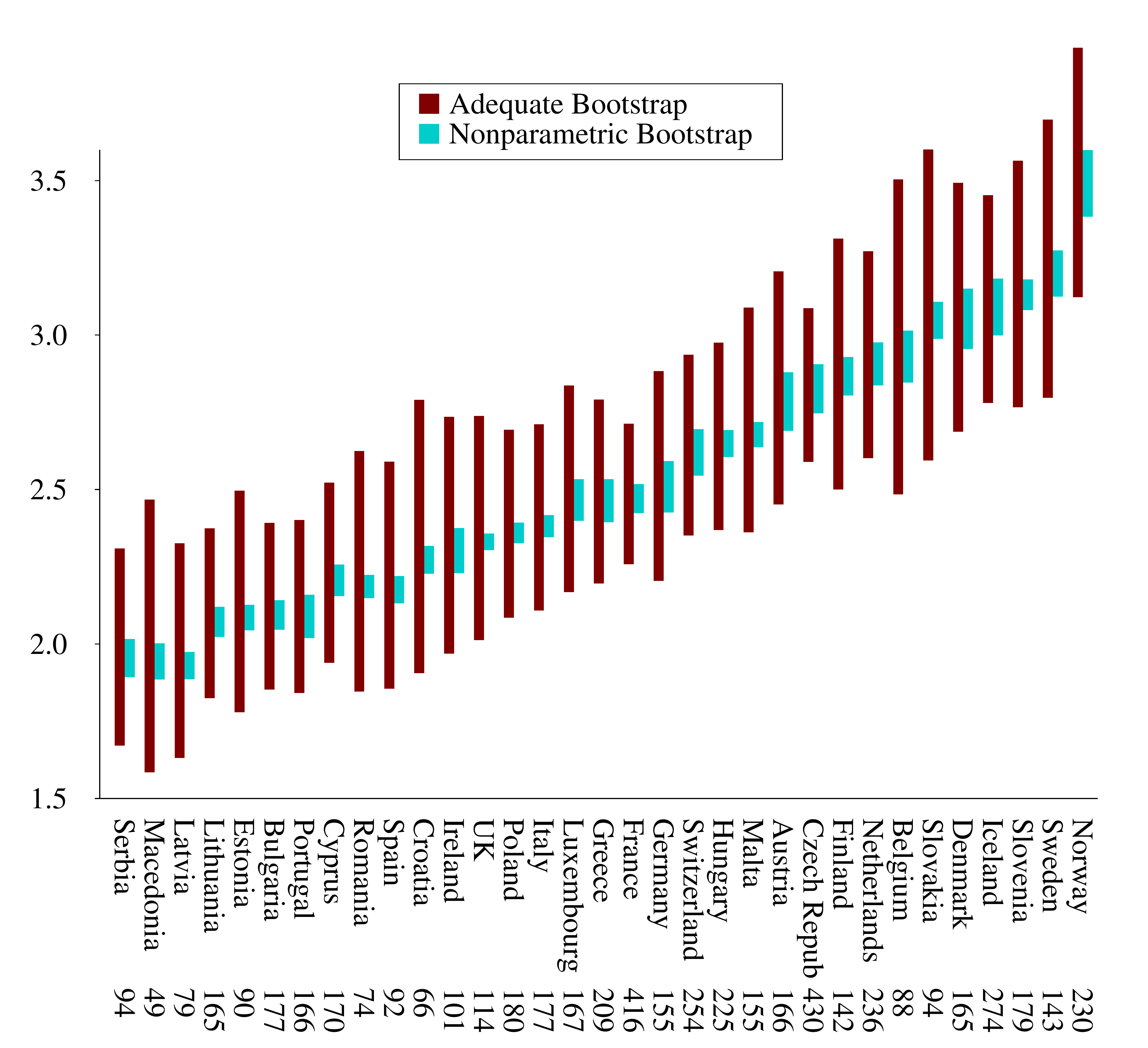}
\caption{Adequate bootstrap confidence intervals for $\alpha$ for a
  Pareto distribution. Adequate bootstrap sizes are listed next to
  country names on the $x$ axis.}\label{IncomePlot}
\end{figure}

Our final real data analysis consists of income data from 33 European
countries. The data are available from
\url{http://ec.europa.eu/eurostat/web/income-and-living-conditions/data/database#}.
A range of percentiles (first to fifth and 95th to 99th percentiles,
and all deciles and quartiles are available) of personal income
distribution are provided for each country. The sample sizes used to
obtain these data are provided in the quality reports at
\url{http://ec.europa.eu/eurostat/c/portal/layout?p_l_id=1012390&p_v_l_s_g_id=0}.
Some of these quality reports are missing, or do not state the sample
size. For our purposes, the exact sample size usually does not matter,
because for the adequate bootstrap, the adequate bootstrap size
depends on the level of fit to the model distribution, not on the
actual sample size. Where information on sample size was not
available, it was assumed to be 5,000. This has had a minor impact on
our results. All sample sizes listed were more than 5,000, so our
choice of 5,000 is conservative, and should lead to wider confidence
intervals. However, the majority of adequate bootstrap sizes were
smaller than 5,000, so it is unlikely that any of these samples would
have had a much higher adequate bootstrap size if the sample size had
been known to be larger. Income distributions are often assumed to follow a
power-law distribution for larger values. We have therefore used a
type-1 Pareto distribution to model the grouped data from the
percentiles, starting at the median (setting the lower bound $\eta$ to
be the median income). The results are shown in
Figure~\ref{IncomePlot}

From the figure, we see that the power law distribution is not a great
fit for these data. In some cases, the adequate bootstrap size is as
little as 49. In some other cases, the adequate bootstrap size is a
little over 400. The adequate bootstrap intervals are all fairly
wide. We should therefore be cautious about drawing conclusions using
a Pareto distribution. The countries are approximately arranged from
lower to higher values of $\alpha$. Using the adequate bootstrap
confidence interval, we see that there are not so many significant
differences in $\alpha$ between countries.

\begin{figure}[htbp]
  \hfil\includegraphics[width=5cm,clip=true,trim=0cm 0cm 0cm 1.5cm]{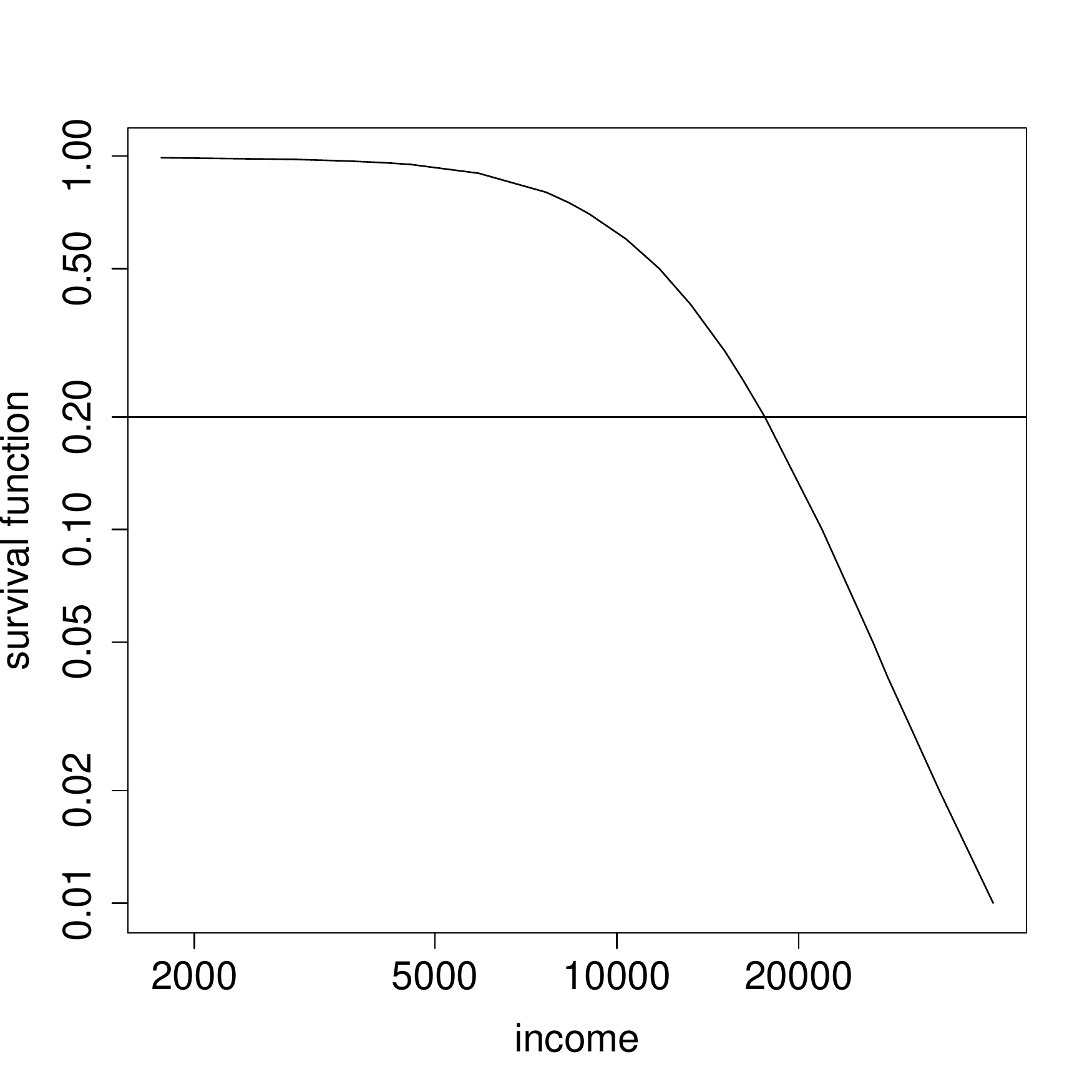}
\caption{Log-Log plot of survival function of income distribution.
  Geometric average of percentiles for 34 countries. On the log-log
  scale, this corresponds to an arithmetic average. Under the
  power-law model, this should be a straight line in the tail of the
  distribution.}\label{IncomeLogLog}
\end{figure}

Looking at a log-log plot of the survival functions (Figure~\ref{IncomeLogLog}), we see that the
problem is that the power law does not apply until higher
percentiles. From the log-log plot, it looks as though the top 20\% of
incomes are close to following a power law.

\begin{figure}[htbp]
  \hfil\includegraphics[width=8cm,clip=true,trim=0cm 1.1cm 0cm 1.4cm]{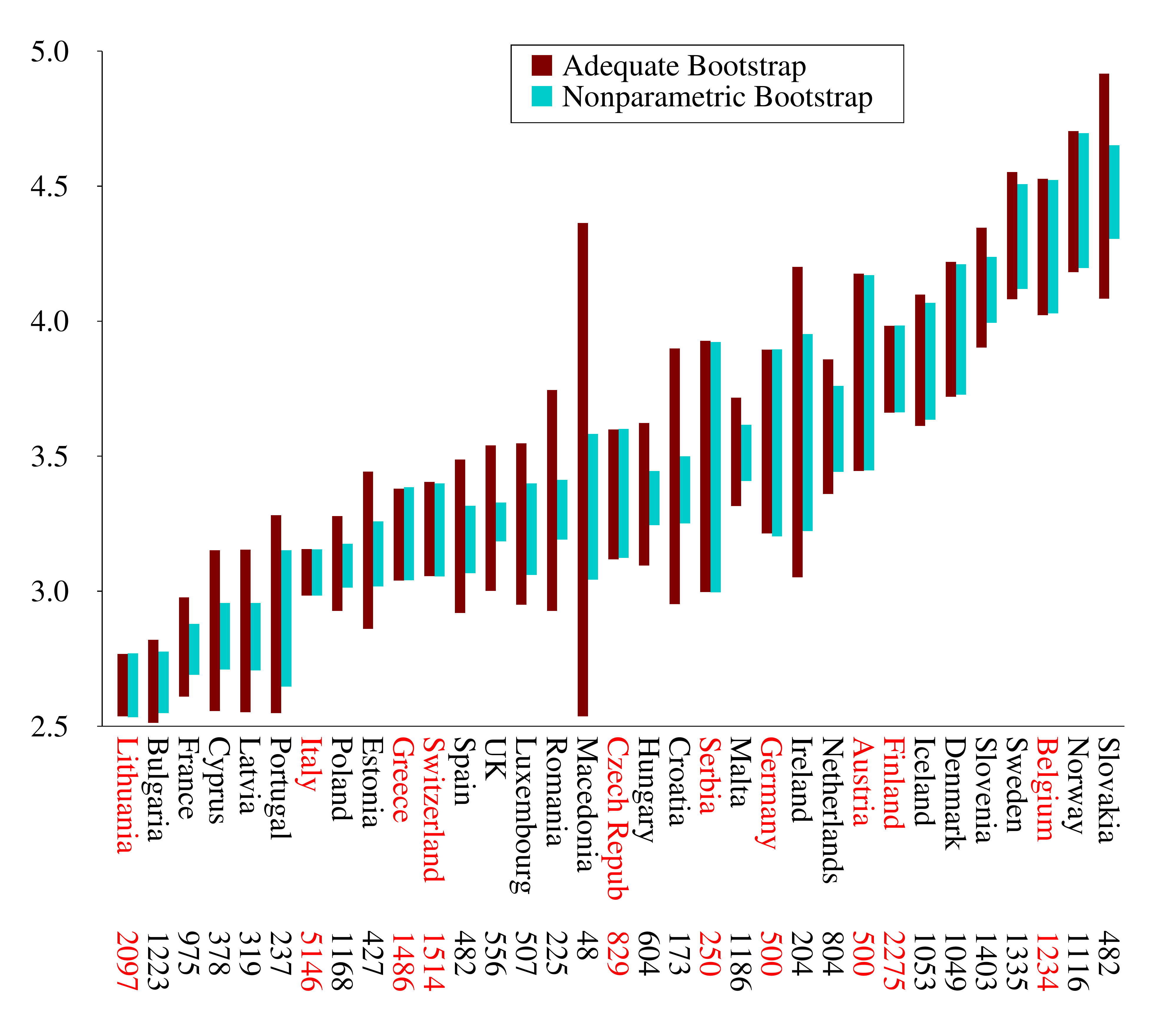}
\caption{Adequate bootstrap confidence intervals for $\alpha$ for a
  Pareto distribution for top 20\% of incomes. Adequate bootstrap
  sizes are listed next to country names on the $x$
  axis. The model was found adequate for countries labelled in red.}\label{IncomePlot20}
\end{figure}

 We therefore calculate the
adequate bootstrap confidence intervals based on the top 20\% of
incomes. This is shown in Figure~\ref{IncomePlot20}.

As indicated by the graphical examination, the Pareto distribution is
a much better fit for this upper tail of the distribution. Indeed it
is deemed adequate in a number of cases. The adequate bootstrap size
has gone up for nearly all cases, and for many cases, there has been a
corresponding decrease in the confidence interval width. Not all
confidence intervals have decreased in width. This can be explained by
the fact that although the adequate bootstrap size has increased, the
data now consists of the upper tail of the distribution, so has fewer
groups, and each data point therefore gives less information about
parameter estimates. 

We note also that the parameter $\alpha$ has increased from the upper
half of the data. This is consistent with the shape of the log-log
survival plots (all concave). It also suggests that indeed a power-law
was not appropriate for the upper half of the data, since if a power
law is appropriate, then the index $\alpha$ should be constant when we
truncate the data. We found for the web links dataset, that the
power-law could be quite a useful model for the data even though it
did not fit the tail well. However, in that case, the tail was a much
more extreme tail with only a very small proportion of data points,
whereas here, the tail is the top 20\% of the original data, or the
top 40\% of the top-half data. We would not expect a power-law to fit
well in cases where the estimated index is very different for the
top-half and bottom half of the index. The ordering of the countries
is fairly consistent between the two plots, with only a few countries
appearing in very different positions on the two plots. It seems to be
the countries for which the power-law fit the data best (France and
the Czech republic) that changed position most in the new
ordering. This makes some sense, since as we observed, the estimated
power-law index increases as we move to the upper tails of the
distribution. For distributions where it increases more, the fit of
the power-law distribution must be less good, whereas for the
countries where the power-law holds for a larger portion of the data,
the estimate should increase less as we move towards the tail. (If the
power-law were correct, we would expect the estimate to remain
approximately constant as we move towards the tail.)

\section{Conclusions and Future Work}\label{CONCLUSION}

\subsection{Conclusions}

The adequate bootstrap is a general method for incorporating model
uncertainty into our inference of parameter estimates. It provides a
more helpful assessment of model adequacy relative to actual needs,
than a simple model adequacy test, which does not take the user's
needs into account at all, and merely tests whether there is enough
data to falsify the model. Lindey and Liu's (2009) credibility index
is a similar attempt to quantify how useful a model is, by estimating
the amount of data that is needed to falsify the model. However, the
credibility index does not give a helpful or intuitive measure of how
well the model fits the data --- most scientists would not be able to
provide a value for an acceptable credibility index, and what is
acceptable will vary between different models. It can also vary with
the data. For example, using grouped data and a chi-square test, the
number of groups used can affect the credibility index much more than
the adequate bootstrap interval. On the other hand, the adequate
bootstrap quantifies the model adequacy in direct terms of the
corresponding variability of parameter values. This is a far easier
measure to assess.

We have shown by means of simulations and theoretical calculations
that in a range of situations where there is a good and natural
interpretation of true parameter values, the adequate bootstrap
provides a good confidence interval with far better coverage than the
usual bootstrap. The width of the confidence interval adjusts with the
adequacy of the model, so a good model will have a narrow confidence
interval, but a less good model will have a wider confidence
interval. Because we do not have a clear underlying model for the
truth, we cannot control the coverage as accurately as we do in
standard parametric inference. We have shown how, in theory, the
coverage can be well controlled in situations where we know the model
used for simulation. However, knowing the model used for simulation is
an unrealistic assumption. If we know in what way the data
distribution should differ from the model distribution, we would do
better to incorporate this into a new parametric or Bayesian model
directly. The adequate bootstrap is intended to be applicable in cases
where the exact form of model misspecification is unknown.

We have provided two examples where this kind of ``wrong'' parametric
approach is appropriate. These examples were chosen both because they
are important examples of the sort of situation where this analysis is
appropriate, and because the theory behind them is relatively
tractible. There are a number of other common situations where the
same philosophy that the parameters of the ``wrong'' model are
meaningful, and an analysis such as the adequate bootstrap is
therefore valuable. We discuss some of these situations, and what
needs to be done to apply the adequate bootstrap technique to them in
the Future Work section. In the examples we gave, some of the
parameters estimated corresponded to non-parametric quantities of the
underlying distribution (the mean and standard deviation). However,
our method did not in any way rely on this fact, and it could be used
just as easily on parameters that have specific meaning in the
specific model. In the power law example, the index of the power law
is not a natural non-parametric quantity, so we could not construct a
non-parametric estimate so easily. It is however a natural quantity
for power law distributions, and can be interpreted even if the
power-law assumption is not perfect.

We have demonstrated the use of this method on three real data
analyses. Here we do not know true parameter values that should be
found, or for the power-law examples, whether there even are
well-defined ``true'' parameter values to estimate. For the stock
market data, we can compare our confidence intervals to
non-parameteric intervals. We see that, even though the fit of the
models is only moderate, in some cases our adequate bootstrap gives a
similar width of confidence interval to the non-parametric method. We
interpret this by saying that because our fitted model is close to the
truth, the estimated confidence interval makes sense in light of our
previous belief that the log-normal model should be approximately
right. The weight we attach to this adequate bootstrap interval might
depend on the strength of our prior beliefs that a log-normal
distribution should be a good model for this data. If we strongly
believe the log-normal model, perhaps with explanations in terms of a
hypothesised data-generating model, then we would attach more weight
to the adequate bootstrap interval. If on the other hand, we had chosen the
log-normal model purely because it offered a good fit to the data at
hand, then we would not put so much weight on the adequate
bootstrap. In the stock market case, a log-normal makes intuitive
sense because the weekly returns are a geometric average of daily and
hourly returns, so under some assumptions, by the central limit
theorem, should be approximately log-normal. We would therefore be
inclined to accept a log-normal as a close approximation to the truth,
and where the data do not offer very strong evidence against the
log-normal, we would be prepared to accept the adequate bootstrap if
it provides a shorter confidence interval.

\subsection{Future Work}

There are a number of areas where the adequate bootstrap can be
further developed. The first is the computational issue. Performing
enough bootstraps to gain a good estimate of the adequate bootstrap
size involves a lot of computation, particularly if the sample size is
large. A lot of work could be saved by improving the method
here. Theorem~\ref{AdBootSize} gives the method for estimating the
adequate bootstrap size from a given set of bootstrap results, but it
does not provide a good method for choosing the bootstraps to perform
for this purpose. A lot more work could go into this area, to provide
a computationally faster technique. The current heuristic algorithm is
reasonably fast for the data sets used, taking something in the order
of 1--2 minutes for smallish sample sizes. However, for more elaborate
models, such as models in phylogeny, estimating a tree and parameter
values from a single dataset can take several minutes or even
hours. To then repeat this procedure for thousands of bootstraps is
completely infeasible. Therefore, a method is needed to obtain the
adequate bootstrap far more efficiently in terms of the number of
bootstraps needed. This is a problem that has been studied in isotonic
regression, and there have been a number of approaches developed. Most
methods for this are based on the ``up and down'' method presented in
\citet{DixonMood}. The idea is to choose the level for each sample
based on the outcome of the previous sample or samples. There has been
substantial work on improving this, e.g. \citep{Narayana},
\citep{DurhamFlournoyRosenburger}, \citep{IvanovaEtAl}. The main
application of isotonic regression is in dose-response trials,
sometimes with human subjects. For such experiments, trials involve
giving subjects doses of a potentially toxic substance. As a
consequence of this, methods are designed based on the principle that
the cost of a ``success'' is greater than that of a failure, so are
designed to minimise the number of successes needed to find the
relevant quantile. In our case, where a trial involves running a
computer program on some data, the cost of each trial is rather closer
to being equal for each sample size (though larger subsample sizes do
often involve more computation). We will therefore want to modify the
methods used in clinical trials to focus more on information gained,
rather than cost.

Another major direction that needs to be developed is how to adapt the
adequate bootstrap to cases where bootstrap size is not the only
factor that determines the probability of passing the adequacy test. A
good example of where this situation arises is in regression. For
regression, the width of a confidence interval, and the significance
of an adequacy test depends not just on the number of points, but also
the influence, or leverage of each point. Points with an $X$-value
near the mean have little leverage, so in a bootstrap where the
$X$-values of the sample have small variance, the model adequacy test
is less likely to be rejected than in a sample where the $X$-values
have large variance. This means that instead of a simple
one-dimensional adequate bootstrap size, we need to consider some
combination of bootstrap size and leverage. Once the appropriate
combination is established, it should be a simple matter to use this
combination in place of the adequate bootstrap size. Censored data is
another example of a situation where points have different influence
--- censored points may be viewed as giving less information than data
points, so our adequate bootstrap would need to be based on some
combination of the number of censored and the number of uncensored
data points.

Finally, a lot more can be done in applying the adequate bootstrap to
new situations. It is generally most appropriate in situations where
the parameters make sense even with a wrong model. Besides the
contamination and sampling bias situations discussed in this paper,
phylogenetic tree estimation is a good example where this technique
makes sense philosophically --- even if our model of the molecular
evolution process is not correct, we still believe that some tree-based
Markov process should apply, so attempts to find an adequate bootstrap
confidence interval for the tree make sense. The idea is that in many
cases, the model will be close enough to the truth to allow us to
confidently infer the true tree. In others, the model will be so
inaccurate, or the tree so uncertain, that the adequate bootstrap
interval will contain many trees, and the adequate bootstrap will show
that the model used is not useful for inferring the phylogenetic tree.
Another situation where this could be used is for prediction under a
regression model. In this case, there are assumptions such as the
linearity of the regression that might not be exactly true. However,
the adequate bootstrap can show how much the uncertainty about this
method would affect the prediction. Finally, a good case where the
adequate bootstrap could be used is in asymptotics. There are many
cases where we are considering a statistic from a finite sample whose
limiting distribution we know. For example, the central limit theorem
tells us the asymptotic distribution of the sample mean; or various
test statistics are known to follow certain asymptotic
distributions. In some cases these asymptotic distributions have
unknown parameters which need to be estimated, and cannot be estimated
well non-parametrically. For example, in extreme value
distributions, it is known that for a wide class of distributions, the
extreme value distribution has a certain parametric form. We therefore
know that the finite-sample distribution is approximately a
distribution of this form, and we may wish to estimate the parameters
of the asymptotic distribution from the finite sample. There will be
some error in this estimation introduced by the fact that we are using
a finite sample distribution instead of the asymptotic
distribution. The adequate bootstrap will tell us how much this might
influence our parameter estimates.

\clearpage

\begin{appendices}

\section{MLE Calculation}\label{AppendixMLE}

Recall that for the sampling bias simulation, the log-likelihood for
this model is
$$-\sum \frac{x_i{}^2}{2\sigma^2}-n\log(\sigma)+\sum_{i=1}^J
n_i\log(p_i)-n\log(\sum T_ip_i)$$
where:
\begin{itemize}
\item $n$ is the sample size

\item $n_i$ is the number of observations in the interval
  $(c_{i-1},c_i]$.

\item $T_i$ is the probability that a random sampling point from a
  normal distribution with mean 0 and variance the current estimate of
  $\sigma$ lies in $(c_{i-1},c_i]$. That is
    $T_i=\Phi\left(\frac{c_i}{\sigma}\right)-\Phi\left(\frac{c_{i-1}}{\sigma}\right)$.
\end{itemize}

Setting the derivative with respect to $p_i$ to zero gives:

\begin{align*}
\frac{n_i}{p_i}-\frac{nT_i}{\sum T_jp_j}&=0\\
p_i&=\frac{\sum T_jp_j}{n}\left(\frac{n_i}{T_i}\right)\\
p_i&\propto\frac{n_i}{T_i}
\end{align*}

Since rescaling all the $p_i$ by a constant does not affect the
distribution, we might as well set $p_i=\frac{n_i}{T_i}$.

Making this substitution, the log-likelihood for a given value of
$\sigma$ is 
$$-\sum \frac{x_i{}^2}{2\sigma^2}-n\log(\sigma)+\sum
n_i\log(n_i)-\sum
n_i\log(T_i)-n\log(\sum n_i)$$
Removing the terms that do not depend on $\sigma$, we get
$$-\sum \frac{x_i{}^2}{2\sigma^2}-n\log(\sigma)-\sum
n_i\log(T_i)$$
The derivative with respect to $\sigma$ is therefore

\begin{align*}
\sum \frac{x_i{}^2}{\sigma^3}-\frac{n}{\sigma}-\sum n_i\frac{\left(\frac{dT_i}{d\sigma}\right)}{T_i}\\
\end{align*}

The second derivative is 

\begin{align*}
-3\sum \frac{x_i{}^2}{\sigma^4}+\frac{n}{\sigma^2}+\sum n_i\left(\frac{\left(\frac{dT_i}{d\sigma}\right)^2}{T_i{}^2}-\frac{\left(\frac{d^2T_i}{d\sigma^2}\right)}{T_i}\right)\\
\end{align*}

We can then find the MLE using Newton's method. (We can take the
sample variance as a starting point.)

\section{Calculation of Hessian Matrix for Sample Bias Simulation}\label{AppendixHessian}

If we let $l$ be the expected
  log-likelihood under the original normal distribution with mean 0,
  variance 1, then this is given by

$$l=-\frac{1}{2\sigma^2}-\log(\sigma)+\sum_{i=1}^J
\pi_i\log(p_i)-\log(\sum_{i=1}^J T_ip_i)$$

Where $\pi_i$ is the probability of a random point lying in the interval
$(c_{i-1},c_i]$ under the true model, so for our case we have
  $\pi_i=\frac{1}{J}$.

The derivative with respect to $p_i$ is
$\frac{\pi_i}{p_i}-\frac{T_i}{\sum T_jp_j}$, while the derivative with
respect to $\sigma$ is $$\frac{1}{\sigma^3}-\frac{1}{\sigma}-\frac{\sum_{i=1}^J p_i\left(\frac{dT_i}{d\sigma}\right)}{\sum_{i=1}^J T_ip_i}$$

The second derivative with respect to $p_i$ and $p_j$ is
$\frac{T_iT_j}{\left(\sum p_kT_k\right)^2}$ for $i\ne j$ and
$\frac{T_i{}^2}{\left(\sum p_kT_k\right)^2}-\frac{\pi_i}{p_i{}^2}$ if
$i=j$. The second derivative with respect to $\sigma$ and $p_i$
is $$-\frac{\left(\frac{dT_i}{d\sigma}\right)}{\sum
  T_jp_j}+\frac{T_i}{\left(\sum
  T_jp_j\right)^2}\sum\left(\frac{dT_j}{d\sigma}\right)p_j$$ and the second derivative with respect to
$\sigma$ is 
$$\frac{1}{\sigma^2}-\frac{3}{\sigma^4}-\left(\frac{\sum
  p_i\frac{d^2T_i}{d\sigma^2}}{\left(\sum
  T_ip_i\right)}-\frac{\left(\sum
  p_i\frac{dT_i}{d\sigma}\right)^2}{\left(\sum
  T_ip_i\right)^2}\right)$$

We also have 
\begin{align*}
\frac{dT_i}{d\sigma}&=\frac{1}{\sqrt{2\pi}}\left(\frac{c_{i-1}}{\sigma^2}e^{-\frac{c_{i-1}{}^2}{2\sigma^2}}-\frac{c_{i}}{\sigma^2}e^{-\frac{c_{i}{}^2}{2\sigma^2}}\right)=\frac{1}{\sqrt{2\pi}\sigma^2}\left(c_{i-1}e^{-\frac{c_{i-1}{}^2}{2\sigma^2}}-c_{i}e^{-\frac{c_{i}{}^2}{2\sigma^2}}\right)\\
\end{align*}

We will evaluate the expected hessian at the ``true'' model $\sigma=1$
and $p_i=1$. This gives $T_i=\frac{1}{J}$. Also note that
$\sum_{i=1}^J p_iT_i=\sum_{i=1}^J T_i=1$ for all values of
$\sigma$. This means that $\sum
  p_i\frac{dT_i}{d\sigma}=\frac{d}{d\sigma}\left(\sum
  T_i\right)=0$ and $\sum
  p_i\frac{d^2T_i}{d\sigma^2}=\frac{d^2}{d\sigma^2}\left(\sum
  T_i\right)=0$. We therefore get  

\begin{align*}
\frac{\partial^2l}{\partial\sigma^2}&=1-3+0-0+0=-2\\
\frac{\partial^2l}{\partial\sigma\partial
  p_i}&=-\left(\frac{dT_i}{d\sigma}\right)-T_i\sum\left(\frac{dT_j}{d\sigma}\right)=-\left(\frac{dT_i}{d\sigma}\right)=-\frac{1}{\sqrt{2\pi}}\left(c_{i-1}e^{-\frac{c_{i-1}{}^2}{2}}-c_{i}e^{-\frac{c_{i}{}^2}{2}}\right)\\
\frac{\partial^2l}{\partial p_i\partial
  p_j}&=\frac{1}{J^2}\qquad\textrm{if }i\ne j\\
\frac{\partial^2l}{\partial p_i{}^2}&=\frac{1}{J^2}-\frac{1}{J}\\
\end{align*}

\section{Additional Tables}\label{AppendixTables}

\begin{table}[htbp]
\caption{Theoretical Coverages}\label{TheoreticalCoverage}

\hfil\begin{tabular}{r|lllllllll}
\hline
\backslashbox {$k$}{$m$} & 1 & 2 & 3 & 4 & 5 & 6 & 7 & 8 & 9 \\
\hline
 1 & 0.500 & 0.375 & 0.311 & 0.272 & 0.245 & 0.225 & 0.209 & 0.197 & 0.187  \\
 2 & 0.625 & 0.500 & 0.426 & 0.376 & 0.339 & 0.311 & 0.289 & 0.271 & 0.256  \\
 3 & 0.689 & 0.574 & 0.500 & 0.447 & 0.406 & 0.374 & 0.349 & 0.327 & 0.309  \\
 4 & 0.728 & 0.624 & 0.553 & 0.500 & 0.458 & 0.425 & 0.397 & 0.373 & 0.353  \\
 5 & 0.755 & 0.661 & 0.594 & 0.542 & 0.500 & 0.466 & 0.437 & 0.412 & 0.391  \\
 6 & 0.775 & 0.689 & 0.626 & 0.575 & 0.534 & 0.500 & 0.471 & 0.445 & 0.423  \\
 7 & 0.791 & 0.711 & 0.651 & 0.603 & 0.563 & 0.529 & 0.500 & 0.474 & 0.452  \\
 8 & 0.803 & 0.729 & 0.673 & 0.627 & 0.588 & 0.555 & 0.526 & 0.500 & 0.477  \\
 9 & 0.813 & 0.744 & 0.691 & 0.647 & 0.609 & 0.577 & 0.548 & 0.523 & 0.500  \\
\hline
\end{tabular}

\end{table}

\begin{sidewaystable}[htbp]
\thisfloatpagestyle{empty}

\caption{Data sets used for power law analysis.}\label{PowerLawDatasetsSources}

\setlength{\tabcolsep}{0.1cm}
\hspace{-1.3cm}\begin{tabular}{lll}
\hline
Name & Source & Reference\\
\hline
SNP &\url{http://siblisresearch.com/data/market-caps-sp-100-us/}
& \\
City & {\tt poweRlaw}
package in {\tt R} & \cite{Arcaute}\\
Country &\url{http://stats.oecd.org/Index.aspx?DatasetCode=IDD} &\\
Moby Dick &{\tt  poweRlaw} package in {\tt R}  & \cite{Newman}\\
Citations &
\url{http://physics.bu.edu/~redner/projects/citation/isi.html} & \cite{Redner}\\
Web Links & \url{http://tuvalu.santafe.edu/~aaronc/powerlaws/data/weblinks.hist}
& \cite{Broder}\\
Quakes & \url{http://tuvalu.santafe.edu/~aaronc/powerlaws/data/quakes.txt}
& \cite{Newman}\\
Income & \multicolumn{2}{l}{\url{http://www.statcan.gc.ca/tables-tableaux/sum-som/l01/cst01/famil105a-eng.htm}}\\
\hline
\end{tabular}

\end{sidewaystable}

\begin{table}[htbp]
\caption{Coverage of adequate jackknife confidence interval for
  various settings of number of intervals $J$ and values of
  $\tau$. Results are out of 1000 simulations.}\label{SampBiasSimResCoverJackknife}

\hfil\begin{tabular}{r|rr}
\hline
\backslashbox{$J$}{$\tau$}   &  0.2 & 0.5 \\
\hline
  3 & 994 & 997 \\
  5 & 956 & 956 \\
  8 & 954 & 959 \\
\hline
\end{tabular}\end{table}

\end{appendices}

\bibliography{references}
\bibliographystyle{apalike}

\end{document}